%% file: BNPCE_OganisianRoyMitra.tex
\documentclass[aoas,preprint]{imsart}

\RequirePackage{amsthm,amsmath,amsfonts,amssymb}
\RequirePackage[authoryear]{natbib}
\RequirePackage{graphicx}
\usepackage{makecell}
\usepackage{multirow}
\usepackage{subcaption}
\usepackage{graphicx}
\usepackage{amssymb}
\usepackage{amsmath}
\usepackage{hyperref}

\DeclareMathOperator*{\argmin}{arg\,min}

\startlocaldefs

\endlocaldefs

\begin{document}

\begin{frontmatter}
\title{Bayesian Nonparametric Cost-Effectiveness Analysis: \\ Causal Inference and Adaptive Subgroup Discovery}
\runtitle{Bayesian Nonparametric Cost-Effectiveness Analysis}

\begin{aug}
\author[A]{\fnms{Arman} \snm{Oganisian} \ead[label=e1,mark]{ } },
\author[A]{\fnms{Nandita} \snm{Mitra}  },
\author[B]{\fnms{Emily M.} \snm{Ko} },
\author[C]{\fnms{Jason A.} \snm{Roy}\ead[label=e2]{Rutgers University}}
\address[A]{Division of Biostatistics \\ Department of Biostatistics, Epidemiology, and Informatics \\ University of Pennsylvania \\ \printead{e1}aoganisi@upenn.edu  }

\address[B]{Division of Gynecologic Oncology \\ Department of Obstetrics and Gynecology  \\University of Pennsylvania Health Systems}
\address[C]{Department of Biostatistics and Epidemiology \\ Rutgers University School of Public Health}

\end{aug}

\begin{abstract}
Cost-effectiveness analyses (CEAs) are at the center of health economic decision making. While these analyses help policy analysts and economists determine coverage, inform policy, and guide resource allocation, they are statistically challenging for several reasons. Cost and effectiveness are correlated and follow complex joint distributions which are difficult to capture parametrically. Effectiveness (often measured as increased survival time) and accumulated cost tends to be right-censored in many real-world applications. Moreover, CEAs are often conducted using observational data with non-random treatment assignment. Policy-relevant causal estimation therefore requires robust confounding control. Finally, current CEA methods do not address cost-effectiveness heterogeneity in a principled way - often presenting population-averaged estimates even though significant effect heterogeneity may exist. Motivated by these challenges, we develop a nonparametric Bayesian model for joint cost-survival distributions in the presence of censoring. Our approach utilizes a joint Enriched Dirichlet Process prior on the covariate effects of cost and survival time, while using a Gamma Process prior on the baseline survival time hazard. Causal CEA estimands, with policy-relevant interpretations, are identified and estimated via a Bayesian nonparametric g-computation procedure. Finally, we outline how the induced clustering of the Enriched Dirichlet Process can be used to adaptively detect presence of subgroups with different cost-effectiveness profiles. We outline an MCMC procedure for full posterior inference and evaluate frequentist properties via simulations. We use our model to assess the cost-efficacy of chemotherapy versus radiation adjuvant therapy for treating endometrial cancer in the SEER-Medicare database.
\end{abstract}


\end{frontmatter}


\newpage
\section{Introduction}

Cost-effectiveness analyses (CEAs) are ubiquitous in public health policy and health economics research, with use-cases ranging from treatment comparison to determining drug coverage and informing policy. However, they remain statistically challenging for several reasons. First, cost and effectiveness are often correlated, with joint distributions typically exhibiting extreme skewness and multimodality. In these settings, parametric models that impose strong distributional, linearity, and additivity assumptions are not tenable. Second, in many cases effectiveness is operationalized as gains in survival time - which is prone to right-censoring if subjects drop out before the end of the study. For such patients, we only observe a lower bound on their survival time and accumulated costs. Third, CEAs are often conducted using observational data which are less expensive and more readily available, but are prone to confounding. Valid estimation of CEA contrasts therefore requires adjustment so that differences in cost-effectiveness due to treatment can be disentangled from differences due to confounders.

Early statistical literature \citep{Lin1997, Lin2000, Lin2003, Bang2000} focused on cost estimation, while assuming efficacy was constant between treatments. Cost estimation alone is challenging due to the pathological nature of costs (censoring, skewness, zero-inflation, etc). Our work enhances this literature by developing a joint model for cost and survival time, rather than solely focusing on cost. Previous work decomposed the joint distribution into a product of a marginal survival time distribution and a cost distribution conditional on survival time. \cite{Huang2002} refer to this as a ``calibration regression'' approach. \cite{Handorf2018} and \cite{Huang2002} approach the modeling from a frequentist point of view. While the former uses a fully parametric approach, the latter uses a semi-parametric approach - making only first and second moment assumptions. \cite{Baio2014} took a fully parametric Bayesian approach to joint modeling that did not allow for full covariate adjustment since the data application of interest was from a randomized trial. In contrast, our Bayesian joint modeling approach makes neither strong distributional assumptions nor functional form (e.g., linearity, additivity) assumptions and allows for covariate adjustment.

\cite{Li2018} took a significant step toward robust causal inference in cost-effectiveness. They formulate causal CEA contrasts in terms of potential outcomes and develop a doubly-robust estimation approach that combines separate conditional mean models for cost and survival with a treatment propensity score model. They show that CEA contrasts can be estimated consistently if either the propensity score or cost/survival regressions are correct. We build on this work in several ways. We also formulate CEA contrasts in terms of potential outcomes - endowing these contrasts with explicitly causal interpretations. However, our modeling approach is fully nonparametric and, therefore, more flexible than the doubly-robust estimator. While, the doubly-robust approach only uses data on uncensored subjects (weighted by the inverse probability of being uncensored) our approach uses data from both censored and uncensored subjects potentially generating efficiency gains. Moreover, our approach is a Bayesian model for the full joint cost-effectiveness distribution - not a weighted combination of separate conditional mean models. This in principle allows for full posterior inference for any function of the joint distribution. Finally, our approach allows for covariate-dependent censoring. Though \cite{Li2018} mention an extension to covariate-dependent censoring, the method proposed and analyzed in their paper relies on randomly censored survival times.

Specifically, our proposed method decomposes the full joint cost-effectiveness distribution into a survival distribution, and a cost model conditional on time. We specify a ``local" parametric cost model and a proportional hazard survival model. A Gamma process (GP) prior is placed on the baseline hazard of the survival time distribution while an enriched Dirichlet process (EDP) prior is placed on the cost and survival covariate effects of the local models, jointly. A key property of the EDP is its induced posterior clustering. The EDP probabilistically partitions the dataset into clusters with similar cost-effectiveness covariate effects and associates different ``local" models with each cluster. Thus, the joint posterior model for cost-effectiveness is an adaptive mixture of locally parametric models. It is adaptive in the sense that the number of clusters need not be pre-specified. More or less clusters are introduced depending on the complexity of the cost-effectiveness distribution.

Our work also advances the literature in Bayesian nonparametric (BNP) causal inference. An array of nonparametric priors have been successfully applied to causal inference problems \citep{Xu2014, Xu2018, Hill2011, Kim2017, Roy2016}. For instance, \cite{Roy2018} use an EDP prior to model joint outcome-covariate distributions and apply the model to causal estimation with missing-at-random covariates. However, modeling of bivariate counterfactual outcomes using the EDP and GP has not been explored. In CEAs, heterogeneity in cost-efficacy is typically either ignored in favor of a single, marginal effect estimate or is explored along pre-defined subgroups (e.g. hispanic males). Methods in the heterogenous treatment effects literature such as Bayesian Additive Regression Trees (BART)-based procedures \citep{hahn2017, henderson2017}  and Causal Forests \citep{Athey2019} are distinct from our approach as they focus on estimating individual-level treatment effects. Moreover, these methods cannot be readily applied to the joint outcome setting with censoring. Instead, we use the induced clustering of the EDP to \textit{propose} subgroups in a probabilistically principled way. We can then describe each subgroup of the joint in terms of its covariate, cost, and efficacy distributions and use these to motivate future, targeted studies. We propose a ``Differential Subgroup Index'' which measures how much of the cost-efficacy heterogeneity is explained by the EDP's partitioning of the joint distribution. This helps us assess the meaningfulness of the clusters.

We begin by providing a brief overview of cost-effectiveness and the desirability of causal estimands. We then present our model along with a Markov Chain Monte Carlo (MCMC) algorithm for posterior inference. We incorporate our model into a g-computation framework for posterior causal effect estimation under specified identification assumptions. Finally, we outline how the induced clustering of the EDP can be used to explore heterogeneity. Simulation studies assessing frequentist properties of our causal effect estimates under various censoring scenarios and generating models are conducted. We end with a cost-effectiveness analysis of chemotherapy and radiation therapy treatments for endometrial cancer using SEER-Medicare claims data.

\section{Overview of Relevant Cost-Effectiveness Contrasts}

In this paper, we consider a binary treatment setting where assignment is indicated by $A\in \{0,1\}$. The goal of CEAs is to characterize the relative cost-effectiveness of these two treatments - necessitating both a cost and efficacy measure. In many settings, the total cost, $Y$, includes all costs accumulated under this treatment - e.g., hospitalization and medication costs incurred due to adverse events.  Moreover, costs are typically measured from the \textit{payer's} perspective, not the patient's perspective. In single-payer systems like that of the United Kingdom, this would be the National Health Service (NHS). For older patients in the United States, as in our data analysis to follow, the payer of interest is typically Medicare. Though lifetime costs is often of interest, many CEAs set a duration for cost accrual (e.g. 2-year costs) due to follow-up constraints. In this paper, we consider a survival time effectiveness measure, $D$. This is the dominant effectiveness measure in cancer CEAs, the motivating data application of our paper.

A typical observational CEA study follows diagnosed patients after assignment to one of two treatment regimes. After some follow-up period, everyone's (possibly censored) cost and survival time, are recorded and various cost-effectiveness contrasts can then be computed. For instance, the incremental cost effectiveness ratio (ICER) is given as $ICER = \frac{E[Y \mid A=1] - E[Y \mid A =0]  }{ E[D \mid A=1] - E[D \mid A =0] } $. This measures the average cost per unit of effectiveness (increase in survival time). We can also define a monetary value under each treatment, $MV(\kappa) = D \kappa - Y$. Here, $\kappa$ is the ``willingness-to-pay'' parameter. It is interpreted as the maximum dollar value the payer is willing to give for a one unit increase in effectiveness. It is considered a fixed, user-specified value. Here, we will suppress notational dependence on $\kappa$ by simply writing $MV$ where there is no ambiguity. A treatment with positive $MV$ suggests that accrued gains in life value, $\kappa D$, are greater than accrued costs. Health economists often assess cost-effectiveness via the average net monetary benefit, $E[NMB]  = E[MV \mid A=1] - E[MV \mid A=0]$, where we have again suppressed dependence of $NMB$ on $\kappa$. This contrast is closely related to $ICER$ and can be interpreted as the average difference in monetary value between treatment groups. Note that average NMB can also be written equivalently as $E[NMB]= (E[D \mid A=1] - E[D \mid A=0])\kappa -  (E[Y \mid A=1] - E[Y \mid A=0])$. This is linear function of $\kappa$ with the efficacy differential as the slope and the cost differential as the intercept. Another related quantity is the Cost Effectiveness Acceptability Curve (CEAC), which is a curve comprised of $P(NMB>0)$ plotted for various $\kappa$. 

However, note that $MV$ and $NMB$ presented above have no causal meaning as treated and untreated subjects may differ systematically in observational studies. This is undesirable because many policy questions are inherently causal with the goal being to estimate the average cost-effectiveness that \textit{would have} accrued had everyone taken a particular treatment, possibly counter to fact. Estimation of $MV$ with causal meaning requires (1) an estimate of the joint distribution of cost and survival time while adjusting for confounders and (2) causal identification assumptions. Even if all relevant confounders are measured and included in the model, misspecification of the adjustment model may yield biased estimates of cost-effectiveness contrasts - motivating the need for robust, nonparametric modeling of the joint. In the following sections we first describe a Bayesian nonparametric model for the joint outcome conditional on confounders and treatment. We then define a causal $NMB$ as the difference in average \textit{potential} monetary value that would have accrued under each treatment. We go on to formulate the identification assumptions required to estimate these causal quantities using our nonparametric joint model.

\section{Joint Nonparametric Model for Cost and Survival Time}

We consider a binary treatment setting in which $n$ patients are assigned to treatment $A_i \in \{0,1\}$ at baseline. Suppose we are interested in contrasting cost-effectiveness over $\tau$ periods (e.g. $\tau=2$ year cost-effectiveness). We observe data $\mathcal{D} = \{ Y_i, T_i, X_i, \delta_i \}_{i=1:n}$ from this study. Here, $X_i=(A_i, L_i)$ is a covariate vector that contains the treatment indicator and a vector of $q$ categorical or continuous pre-treatment confounders, $L_i$. For notational convenience, we proceed without an intercept, but note that a $1$ can be included in the first entry of $X_i$. We let $T_i = min(D_i, C_i, \tau)$ be the observed time under study (the minimum of a random right-censoring time $C_i$, end of study $\tau$, and death time $D_i$). Define a censoring indicator as $\delta_i=I(D_i > min(C_i, \tau))$ Finally, $Y_i \in \mathcal{Y}$ denotes cost accumulated through time $T_i$. The joint distribution can be factored into a distribution for observed time and cost distribution conditional on time. A joint model follows from specifying ``local" models for each of these two distributions:
\begin{equation} \label{eq:cemod}
\begin{split}
Y_i \mid T_i, \delta_i, X_i, \omega_i \sim &\ p(Y_i \mid T_i,  \delta_i, X_i, \omega_i) \\
T_i \mid \delta_i, X_i, \theta_i, \lambda_0 \sim & \ \lambda_0(t) \exp( X_i' \theta_i ) \\
\omega_i, \theta_i \mid G  \sim & \ G. \\
\end{split}
\end{equation}
At a particular time, $T$, cost follows some local distribution $p(Y_i \mid T_i, \delta_i, X_i, \omega_i)$ governed by parameters $\omega_i$. Survival time follows some local hazard function which is parameterized as having some baseline hazard, $\lambda_0$ with covariate effects, $\theta_i$, multiplying this baseline hazard. Lastly, $\omega_i$ and $\theta_i$ - the covariate effects of the cost and effectiveness model - both follow some joint prior distribution $G$, which is unknown. Choice of the local models are application-specific but are not crucial for model fit, as will become apparent when we discuss the nonparametric priors used for $G$ and $\lambda_0$. 

One consideration when choosing the local model is desired predictive support. For instance, if costs are sufficiently far from zero, we may be willing to set $p(Y_i \mid T_i, \delta_i, X_i, \omega_i)$ to a Gaussian over $\mathcal{Y} = \mathbb{R}$ with mean and variance $\omega_i = (\mu_i, \phi_i)$. The corresponding regression could be specified as $\mu_i = (T_i, \delta_i, X_i)'\beta_i$. If the non-negative nature of costs must be respected, we could instead specify a log-normal distribution over $\mathcal{Y} = \mathbb{R}^+$. For applications with zero-inflated costs, we may wish to explicitly put positive measure on zero -  i.e. setting $\mathcal{Y} = \{0 \}\cup\mathbb{R}^+$. This can be done by specifying a two-part model $Y_i \mid T_i, \delta_i, X_i, \omega_i \sim \pi_i \delta_0(Y_i) + (1-\pi_i) f(Y_i \mid T_i, \delta_i, X_i, \beta_i )$, where $\pi_i = P(Y_i=0 \mid T_i,\delta_i, X_i, \gamma_i  )$ is a covariate-dependent model for the probability of cost being zero (e.g. a local logistic regression) and $\delta_0$ is the point mass distribution at 0. In this case, the cost parameter vector is $\omega_i = (\gamma_i, \beta_i)$. \cite{oganisian2020} developed a nonparametric Bayesian estimation procedure for such a two-part model, where $f$ could be either log-Normal or Normal, using a Dirichlet Process prior. 

In \eqref{eq:cemod}, censored patients contribute to the likelihood through both the cost and survival time models. In the survival model, they contribute to the likelihood through the survival function in the usual way, provided that, conditional on covariates, censoring times are independent of survival times. In the cost model, dead patients provide information about the cost distribution at death time $p(Y_i \mid T= D_i, \delta =0, X_i, \omega_i)$, while censored subjects inform the model at time of censoring $p(Y_i \mid T=C_i, \delta =1, X_i, \omega_i)$.

\subsection{Nonparametric Priors}

We specify the following nonparametric priors on the unknown model quantities, $G$ and $\lambda_0$.
\begin{equation}
\begin{split}
G \mid \alpha_\omega, \alpha_\theta  \sim & \  EDP(\alpha_\omega,\alpha_\theta, G_0) \\
\lambda_0 \mid b, \lambda_0^*, \xi \sim & GP( b \lambda_0^*, b, \xi),
\end{split}
\end{equation}

\begin{figure}
    \centering
    \begin{subfigure}[t]{0.48\textwidth}
        \centering
        \includegraphics[width=\linewidth]{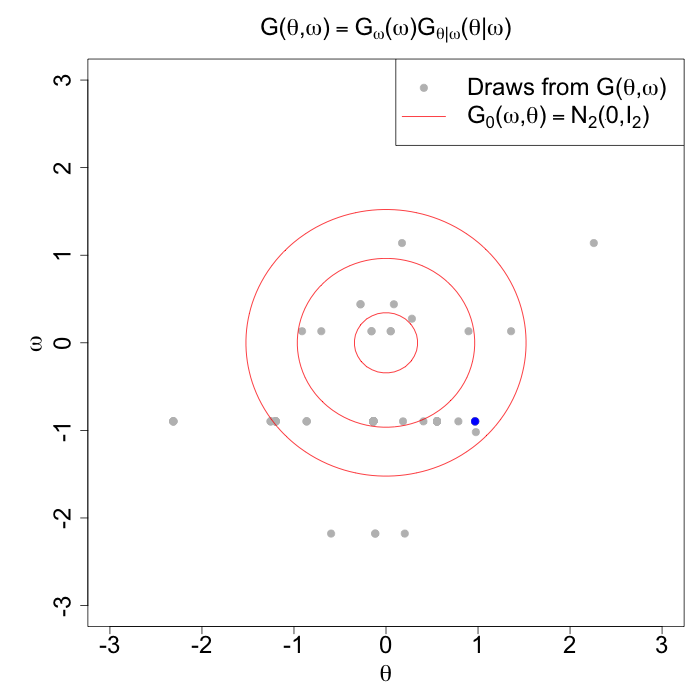} 
        \caption{} 
        \label{fig:edpdraws}
    \end{subfigure}
    \hfill
    \begin{subfigure}[t]{0.48\textwidth}
        \centering
        \includegraphics[width=\linewidth]{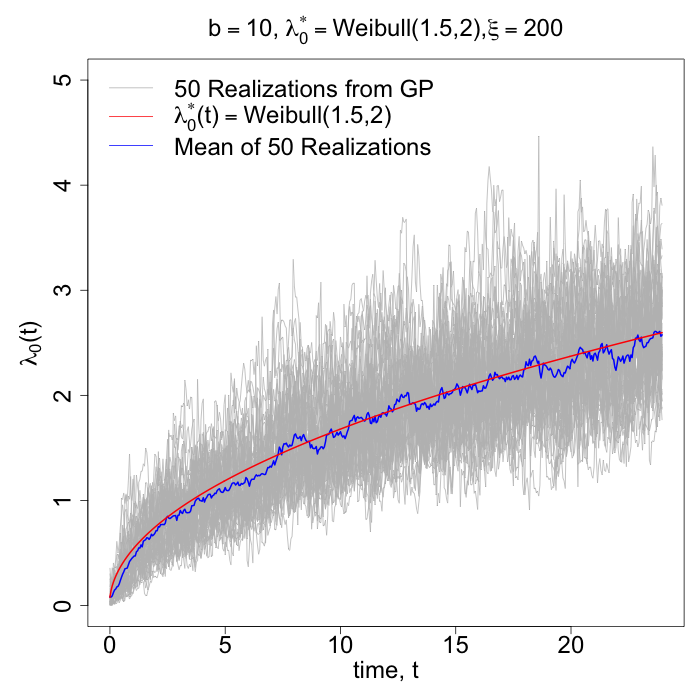} 
        \caption{} 
        \label{fig:gpdraws}
    \end{subfigure}
    \vspace{.1in}
    \caption{Realizations of the Enriched Dirichlet and Gamma Processes. (a) 100 draws of $(\theta, \omega) \sim G$ where $G\sim EDP(10,10, N_2(0, I_2))$. Note the nested discreteness of $G$ causes ties (i.e. clustering) among the draws: there are 80 other draws with the same $\omega$ value as the blue point, but with different $\theta$ values. Twenty three of those 80 also have the same $\theta$ value. (b) Gray lines show 50 hazard realizations from a gamma process centered around the hazard of a $Weibull(1.5, 2)$ distribution. The blue line shows the mean of the 50 realizations. }
\end{figure}

Above, EDP denotes the Enriched Dirichlet Process \citep{Wade2014} prior on $G$ and GP denotes the dependent Gamma Process prior \citep{barajas2002} on the baseline hazard $\lambda_0$. These priors are nonparametric in the sense that they are probability measures on infinite-dimensional objects - the former over probability distributions and the latter over hazard functions. Realizations, $G$, from the EDP are discrete probability distributions centered around a base distribution $G_0(\omega_i, \theta_i) = G_{0\omega}(\omega_i) G_{0\theta|\omega}(\theta_i | \omega_i)$ with two concentration parameters, $\alpha_\omega$ and $\alpha_\theta$. Some prior realizations are visualized in Figure \ref{fig:edpdraws}. Just as with the Dirichlet Process (DP), this discreteness induces a posterior clustering of patients. Unlike the DP, the clustering induced by the EDP is nested. \textit{A posteriori}, patients with similar cost parameters are clustered together into what we call $\omega$-clusters. Within each $\omega$-cluster, patients with similar effectiveness parameters are clustered together ($\theta$-clusters). The EDP prior does not require pre-specification of the number of clusters. The clustering is data-adaptive, with more clusters being introduced to capture more complex cost-effectiveness distribution. The posterior model for the joint distribution is an adaptive nested mixture of cost-effectiveness models - with each component model having the form of the local model in \eqref{eq:cemod}, but with different component-specific parameters. In the machine learning literature, these models are often referred to as ``mixture of experts'' learners: the data space are partitioned into homogenous regions, each having its own model that develops ``expertise'' in that region. This is in contrast to ensemble learners (e.g. BART and Random Forests), which apply multiple models to the \textit{entire} data and combine the results post-hoc.

The GP can be thought of as a prior over the space of hazard functions. Each realization $\lambda_0$ from the GP is a hazard function centered around a mean function $\lambda_0^*$ with concentration parameter $b$. Some prior realizations are visualized in Figure \ref{fig:gpdraws}. The process is ``dependent'' in that it induces a prior AR(1) autocorrelation structure on $\lambda_0$: the hazard at time point $t$ is a weighted average of the hazard at the previous time point and the prior hazard, $\lambda_0$. The resulting shrinkage/smoothness, controlled by hyperparameter $\xi$, regularizes the empirical estimate of the baseline hazard - which can be erratic at later time points when the at-risk set becomes small.

These prior choices are motivated by the shortcomings of the standard DP. A potential issue with specifying $G \sim DP(\alpha G_0)$ is that it imposes a single layer of clustering for both cost and effectiveness. Many clusters may be introduced to fit the joint of $Y$ and $T$ if one of these dimensions is more complex - even if the other is very simple. This makes estimates needlessly variable. The nested nature of the EDP avoids this by allowing varying number of clusters on each dimension controlled by separate concentration parameters. Thus, it is possible to introduce a single cost cluster that has many survival time subclusters. Similarly, modeling the baseline hazard separately avoids introduction of excess clusters to fit a potentially complicated function which, for causal estimation purposes, is just a nuisance parameter. This is also the reason why we opt for a proportional hazard (PH) formulation rather than an accelerated failure time (AFT) approach: PH models clearly separate the covariate effects from the baseline risk, which we do not want influencing the EDP mixture.

\subsection{Posterior Inference using Markov Chain Monte Carlo} \label{sc:mcmc}

Inference for \eqref{eq:cemod} is done via MCMC. We follow the general scheme of Neal's algorithm 8 \citep{Neal2000}, which introduces auxiliary parameters to sample from the DP posteriors. \cite{Roy2018} used this approach to sample EDP posteriors, though without a Gamma Process update and no joint outcome considerations. The idea is to introduce latent cluster indicators (the auxiliary parameters) for each subject. Conditional on draws in the previous iteration, each MCMC iteration then updates clustering indicators conditional on parameters and before updating cluster-specific parameters conditional on these newly updated indicators. At iteration $m$, we may have $J^{(m)}$ occupied $\omega$-clusters indexed by $j\in \{1,\dots, J^{(m)}\}$ and, within the $j^{th}$ $\omega$-cluster, we may have $K_j^{(m)}$ occupied $\theta$-clusters indexed by $k_j \in \{1,\dots, K_j^{(m)}\}$. Let $c_{1:n} = (c_1, \dots, c_n)$ be cluster assignment indicators where each $c_i$ is a length two vector with first and second entry indicating membership to an $\omega$-cluster and $\theta$-subcluster, respectively. Throughout, we use the notation $v_{a:b}$, where $a<b$ are integers, to denote the collection $(v_a, v_{a+1},  \dots, v_b)$ . Let $\omega_{[j]}$ represent the cost parameter associated with cluster $j$ and $\theta_{[j,k]}$ represent the effectiveness parameter associated with the $k^{th}$ subcluster of $\omega$-cluster $j$. We should strictly denote $\theta_{[j,k]}$ as $\theta_{[j,k_j]}$ but suppress the subscript throughout wherever reference is clearly made to the $k^{th}$ subcluster of $\omega$-cluster $j$. Moreover, define $n_j^{-i}$ and $n_{j,k}^{-i}$ as the number of subjects (excluding subject $i$) currently occupying $\omega$-cluster $j$ and $\omega$-$\theta$ cluster $(j,k)$, respectively, at the current iteration, $m$. At each iteration $m$ we conduct the following sequence of conditional posterior updates:
\begin{itemize}
	\item \textit{Update cluster membership}: 
	\begin{itemize}
		\item Propose parameters for a new $\theta$-subcluster for each existing $\omega$-cluster, $\{ \theta_{[j,K_j^{(m)}+1]} \ : j \in 1,\dots, J^{(m)} \}$ by drawing from the prior $G_0$.
		\item Similarly, propose parameters for a new $\omega$-cluster with a $\theta$ subcluster, $\{ \omega_{[J^{(m)}+1]}, \theta_{[J^{(m)}+1, 1]} \}$.
		\item Conditional on current draws of all cost-effectiveness parameters and $\lambda^{(m)}_0$ (indicated by ``$-$'' for compactness), update $c_i^{(m)}$ according to the following probabilities:
		\small
		  \[ P(c_i^{(m+1)}=(j,k) \mid -, \mathcal{D}  ) \propto \left\{
                		\begin{array}{ll} 
				\frac{n_j^{-i}n_{j,k}^{-i} }{n_j^{-i} + \alpha_\theta } p(Y_i, T_i \mid X_i, \delta_i, \omega_{[j]}^{(m)}, \theta_{[j, k]}^{(m)}, \lambda^{(m)}_0 ) & \textnormal{for existing $j,k$} \\
				\frac{n_j^{-i} \alpha_\theta }{n_j^{-i} + \alpha_\theta } p(Y_i, T_i \mid X_i, \delta_i, \omega_{[j]}^{(m)}, \theta_{[j, K_j^{(m)}+1]}^{(m)}, \lambda^{(m)}_0  )& \textnormal{for existing $j$, new $k$} \\
				\alpha_\omega  p(Y_i, T_i \mid X_i,\delta_i, \omega_{[J^{(m)}+1 ]}^{(m)}, \theta_{[J^{(m)}+1, K_j^{(m)}+1]}^{(m)}, \lambda^{(m)}_0 ) & \textnormal{new $j,k$}
                		\end{array}
              		\right.
		 \]
	\end{itemize}
	\item \textit{Update cluster parameters}: These require Metropolis-Hastings steps if $G_0\omega$ or $G_{0\theta|\omega}$ are not conjugate.
	\begin{itemize}
		\item Update each cluster's cost parameter, $\omega_{[j]}$, by drawing from conditional posterior 
		\[ \omega_{[j]}^{(m+1)} \sim p(\omega_{[j]} | c_{1:n}^{(m+1)}, \mathcal{D}) \propto G_{0\omega}(\omega_{[j]}) \prod_{i \mid c_i^{(m+1)} \in (j, \cdot)} p(Y_i | T_i, X_i, \delta_i, \omega_{[j]})  \]
		\item For each $j$, update all $\theta_{[j,k_j]}$ by drawing from conditional posterior 
		\[ \theta_{[j,k]}^{(m+1)} \sim p(\theta_{[j,k]} | c_{1:n}^{(m)}, \lambda_0^{(m)}, \mathcal{D}) \propto G_{0\theta|\omega}(\theta_{[j,k]}) \prod_{i \mid c_i^{(m)} \in (j, k)} p(T_i \mid X_i, \delta_i, \lambda_0^{(m)} , \theta_{[j,k]})  \]
	\end{itemize}
	\item \textit{Update baseline hazard, $\lambda^{(m+1)}_0$}: This is a multi-step update involving a discretization of the time interval $[0,\tau]$ into increments, then modeling the hazard rate in each increment. This is motivated by the fact that if $\lambda_0$ follows a Gamma Process, then the hazard rates in any finite partition of the time interval have Gamma distributions \citep{barajas2002}. Additionally, the latent parameters inducing the AR(1) smoothness across increments are also updated with a mix of grid sampling and adaptive Metropolis steps. Details are provided in Appendix \ref{ap:mcmc}.
\end{itemize}

Note that the induced nested clustering of the EDP is explicitly encoded into this sampler. In the cluster-update step, a given subject is most likely to be assigned to the cluster with parameters that yield the highest joint-distribution evaluation (i.e. fit their data the best). Moreover, each subject can possibly be assigned to a new cost cluster, new effectiveness cluster within an existing cost cluster, or a new cost-effectiveness cluster. This last event is likely to occur if, for example, the subject is so unique that random parameter draws from the prior fit that subject's data better than any of the existing cluster-specific parameters. Furthermore, note that each term for an existing cluster in $P(c_i^{(m+1)}=(j,k) \mid -, D)$ is an increasing function of the number of patients already assigned to that cluster. This is the ``rich-get-richer'' property of the EDP - the \textit{a priori} favoring of assignment to larger clusters. This prevents over-fitting by penalizing small clusters. After every cycle, $c_i^{(m+1)}$ maps each subject to a set of updated parameters $(\omega_i^{(m+1)}, \theta_i^{(m+1)}, \lambda_0^{(m+1)})$. After a sufficient burn-in period this algorithm produces $M$ draws from the posterior $\{\omega_{1:n}^{(m)}, \theta_{1:n}^{(m)}, \lambda_0^{(m)}, c_{1:n}^{(m)} \}_{1:M}$. These can be used to do full posterior inference on any functional of the joint including, as we will see, causal estimands.

\subsection{Priors and Hyperparameter Choice}
\label{sc:hyperparms}
The hyperparameters for the EDP are the base distribution $G_0(\omega_i, \theta_i) = G_{0\omega}(\omega_i) G_{0\theta|\omega}(\theta_i | \omega_i)$ and the concentration parameters $\alpha_\theta$ and $\alpha_\omega$. Following previous papers \citep{oganisian2020, Roy2018}, we use prior independence so that $G_0(\omega_i, \theta_i) = G_{0\omega}(\omega_i) G_{0\theta}(\theta_i)$ and set $G_{0\theta}(\theta_i) = N( \hat\theta_{PH}, \nu_{\theta} \hat C_{PH})$. Here, we are centering the cluster-specific covariate effects around the Cox proportional hazard estimate, $\hat \theta_{PH}$. The prior covariance matrix, $\hat C_{PH}$ , is diagonal with the square of the Cox proportional hazard standard error estimates along the diagonal. The parameter $\nu_{\theta}>0$ is a user-specified scalar that controls how tightly or widely dispersed the cluster-specific effects are around the Cox estimates.

The choice of $G_{0\omega}(\omega_i)$ depends on the choice of local cost model. Suppose our local model, $p(Y_i T_i, \delta_i, X_i,  \omega_i)$ is Gaussian, $N(\mu_i, \phi_i)$ with regression $\mu_i = E[Y_i \mid T_i, \delta_i, X_i, \omega_i] = (\delta_i, T_i, X_i)' \beta_i$ and variance $\phi_i$, where $\beta_i$ is the vector of covariate effects. The full cost parameter vector is $\omega_i = (\beta_i, \phi_i)$ and we could set $G_{0\omega}(\beta_i, \phi_i) = N(\beta_i; \hat\beta, \nu_{\omega} \hat \Sigma) IG(\phi_i; shape=a_0, scale= \hat s^2 (a_0 - 1) $. The vector $\hat \beta$ is the MLE estimate of the cost regression and $\hat \Sigma$ is a diagonal matrix with the square of the standard error estimates along the diagonal. The parameter $\nu_\omega>0$ is user-specified and controls the tightness of the prior around $\hat \beta$. Similarly, the Inverse Gamma prior for $\phi_i$ having mean equal to the empirical outcome variance, $\hat s^2 = \frac{1}{n-1} (Y_i - \bar Y)^2$. The user-specified parameter, $a_0$, controls how widely the cluster-specific variances are dispersed around the empirical variance, with higher values corresponding to a tight prior around the empirical estimate. Finally, we follow previous approaches \citep{Roy2018, oganisian2020} and set $Gam(1,1)$ (i.e. flat, uninformative) priors on each of the concentration parameters. These parameters can be interpreted as prior sample sizes for the cost and effectiveness clusters - higher values on average lead to more occupied clustering. Thus, this Gamma prior penalizes many occupied clusters, but has a long tail to allow posterior deviations if demanded by the data.

Finally, we center the Gamma Process prior around a constant hazard function. Specifically, we compute the Nelson-Aalen estimate of the baseline cumulative hazard, then take the difference between each point on this curve to obtain the baseline hazard estimate at each time point. We then compute the average of these hazard rates across time, $\hat \lambda$. Then, in $GP(b\lambda_0^*, b, \xi)$ we can set $\lambda_0^*$ to be exponential with rate $\hat \lambda$. Intuitively, this expresses the prior belief of a constant hazard (with rate in the range of the observed rates). However, if the data disagrees, the posterior will move us to a richer estimate governed by the data. The parameters $\xi$ and $b$ can be used to calibrate degrees of informativeness. For example $\xi$ near zero and large $b$ corresponds to an informative prior belief of a constant hazard. Conversely, values of $b$ near 0 correspond to an uninformative prior.

\section{Posterior Causal Estimation via g-Computation} \label{sc:causal}
Here we describe full posterior inference for various causal estimands expressed in terms of potential outcomes \citep{Rubin1978}. In scenarios with censored outcomes, causal estimands are typically formulated under a hypothetical ``joint intervention'' \citep{robins2000} on \textit{both} treatment and censoring. Let $MV^{A=a,\delta=0} = D^{A=a,\delta=0} \kappa - Y^{A=a,\delta=0}$ be the monetary value that would have accrued over $\tau$ periods had the patient received treatment $a$ and not been censored. The components $D^{a,0}$  and $Y^{a,0}$ are the survival time and costs, respectively, that would have been observed under treatment $A=a$ had the subject not been censored. The \textit{population-level} estimand of interest is $ \Psi = E[NMB] = E[MV^{1, 0 }] - E[MV^{0,0} ]$. This is the average difference in monetary value that would have accrued over $\tau$ periods had everyone in the target population been assigned to treatment 1 versus treatment 0, and not been censored. In general, interventions in observational CEAs are not random. Instead, they are driven by confounders - factors which both influence treatment and cost-effectiveness. Thus, $E[MV^{a,0}] \neq E[MV \mid A=a, \delta=0]$ in general, since those who actually received treatment and remained uncensored may not be representative of the target population. Suppose, however, that we observe a set of pre-treatment confounders, $L$. Under the following extensions of the usual causal identification assumptions, we can identify $\Psi$:
\begin{itemize}
	\item[IA.1] \textit{Joint ignorability}: $ (Y^{a,\delta}, D^{a,\delta}) \perp (A, \delta) \mid L $. Conditional on $L$, censoring and treatment should be as good as random - being completely independent of the death and costs that would have accrued under a particular treatment. Omission of unmeasured drivers of both the joint intervention or cost-effectiveness would result in a violation of this assumption.
	\item[IA.2] \textit{Joint Consistency}: $(Y^{a,0}, D^{a, 0}) = ( Y, D ) \mid A=a, \delta=0$. This requires that cost and death time observed for an uncensored ($\delta=0$) subject assigned treatment $A=a$ is actually $(Y^{a,0}, D^{a, 0})$. This could be be violated if, for instance, we had non-compliance to the treatment. Then, a subject assigned $a$ may not have actually taken $a$ and thus we would not observe $Y^{a,0}$.
	\item[IA.3] \textit{Joint Positivity}:  $0 < P(A=a, \delta=0 \mid L ) < 1$. The joint intervention cannot be deterministic at any level of $L$. This could be violated if, for example, all uncensored males received treatment $A=1$ - leaving us with no information on how well uncensored males with treatment $A=0$ faired. In these cases, the model may extrapolate the outcome under treatment $A=0$ learned from females onto males. Poor extrapolation could lead to bias.
	\item[IA.4] \textit{No Joint Interference}:  $(Y_i^{a_{1:n}, \delta_{1:n}},  D_i^{a_{1:n}, \delta_{1:n}} ) = (Y_i^{a_{i}, \delta_{i}},  D_i^{a_{i}, \delta_{i}} )$. Here, $a_{1:n}$ and $\delta_{1:n}$ are $n-$dimensional vectors containing each subject's treatment and censoring status. This assumption requires that one person's joint treatment-censoring intervention cannot impact another's cost-effectiveness. It allows us to drop all but the $i^{th}$ element of $a_{1:n}$ and $\delta_{1:n}$. Usually this assumption would be violated in infectious disease exposures or other settings where subjects cannot be reasonably viewed as exchangeable (one person's infection status may impact another's infection probability).
\end{itemize}

Under these assumptions, $\Psi$ is identified via Robins' g-formula \citep{Robins1986}
\begin{equation} \label{eq:mcmv}
\begin{split}
	\Psi( \omega_{1:n}, \theta_{1:n}, \lambda_0) & = \int_{\mathcal{L}} \Big( E[MV \mid A=1, \delta=0, L,  \omega_{1:n}, \theta_{1:n}, \lambda_0] \\ 
			& \ \ \ \ \ \ \ \  - E[MV \mid A=0, \delta=0, L,  \omega_{1:n}, \theta_{1:n}, \lambda_0]  \Big)dP(L) 
	\end{split}
\end{equation}
Details are provided in the Appendix \ref{ap:identification}. Above, we have explicitly written $\Psi = \Psi( \omega_{1:n}, \theta_{1:n}, \lambda_0)$ as a function of the parameters governing the joint cost-effectiveness distribution. This is to highlight that a posterior distribution over these parameters induces a posterior on the the causal estimand $\Psi$. Let each expectation in \eqref{eq:mcmv} be denoted as $\mu(a, 0)=E[MV \mid A=a, \delta= 0, L,  \omega_{1:n}, \theta_{1:n}, \lambda_0]$. Then, 

\begin{equation} \label{eq:expmv}
	\mu(a,0) = \int_0^\tau \int _0^\infty (D \kappa - Y) p(Y, T \mid L, A=a, \delta=0, \omega_{1:n}, \theta_{1:n}, \lambda_0 ) dY dD \
\end{equation}
Where this inner integration is over the joint model we presented in \eqref{eq:cemod} with $X_i = (A_i, L_i)$. Note that conditional on $\delta=0$, $T=D$ in the joint model and we integrate along the time up until $\tau$ - resulting in $\tau$-period monetary value. This integration can be done efficiently via Monte Carlo (see Appendix \ref{ap:mcmc}). 

The outer integration over $\mathcal{L}$ in \eqref{eq:mcmv} requires an estimate of $P(L)$. To avoid strong parametric assumptions, we use a Bayesian bootstrap \citep{Rubin1981}. That is, we express $p(L)$ as a discrete distribution with mass $p_i$ at the $i^{th}$ observed confounder vector $L_i$. Specifically, $p(L = l) = \sum_{i=1}^n p_i \cdot \delta_{L_i}(l) $. Here $\delta_{L_i}$ is a point-mass at $L_i$. The Bayesian bootstrap follows from an improper Dirichlet prior on the weights, $ p_{1:n}=(p_1, \dots, p_n) \sim Dir(0,\dots, 0)$. This yields a conjugate posterior $p_{1:n} \mid L \sim Dir(1, \dots, 1)$ with $n-$dimensional posterior mean vector $E[p_{1:n}\mid L ] = (1/n, 1/n, \dots, 1/n)$.

At the end of the $m^{th}$ iteration of updates from Section \ref{sc:mcmc}, we have a set of parameter draws $\{ \omega_{1:n}^{(m)}, \theta_{1:n}^{(m)}, \lambda_0^{(m)} \}$, which we can use to construct a posterior draw of monetary value $\mu_i^{(m)}(a, 0)=E[MV \mid A=a, \delta= 0, L_i,  \omega_{i}^{(m)}, \theta_{i}^{(m)}, \lambda_0^{(m)}]$. We then take a draw $p_{1:n}^{(m)}$ from the Dirichlet posterior and construct a draw of the confounder distribution $p^{(m)}(L = l) = \sum_{i=1}^n p_i^{(m)} \cdot \delta_{L_i}(l) $. Substituting both of these into \eqref{eq:mcmv}, yields a draw from the posterior of $\Psi$

\begin{equation} \label{eq:mcpsi}
	\Psi^{(m)} \approx \sum_{i=1}^n  p_i^{(m)} \Big( \mu^{(m)}_i(1, 0) - \mu^{(m)}_i(0,0)   \Big)
\end{equation}

Repeating for iterations $m=1, \dots, M$ yields $M$ draws from the posterior of the causal $\tau$-period NMB: $\{ \Psi^{(m)}  \}_{1:M}$. The mean of these draws can serve as a Bayesian nonparametric point estimate of $\Psi$ and percentiles of the $M$ draws can be used to form credible intervals.

The posterior draws can also be used to compute a point on the CEAC for each $\kappa$, $P(NMB >0 \mid D) \approx \frac{1}{M} \sum_m I(\Psi^{(m)} >0)$. We note that, from this Bayesian perspective, each point on the CEAC is a posterior p-value or tail-area probability. If individual-specific estimates are required, Equation \eqref{eq:expmv} can be evaluated for particular $L_i$ under both treatments using each of the $m$ posterior parameter draws. The difference would be a draw from the posterior of $\Psi_i = NMB_i(\kappa)$, denoted $\Psi_i^{(m)}= \mu^{(m)}_i(1, 0) - \mu^{(m)}_i(0,0) $. In the causal literature, these are variously referred to as conditional average treatment effects (CATEs) or individual treatment effects (ITEs). Across $M$ iterations, we would also have subject-level credible intervals for $\Psi_i$. Figure \ref{fig:indnmb} visualizes posterior mean and intervals for each $\Psi_i$ using an illustrative synthetic example.

\section{Adaptive Subgroup Discovery} \label{sc:subgroup}
The MCMC scheme of Section \ref{sc:mcmc} yields posterior draws of latent cost-effectiveness cluster membership, $\{ c_i^{(m)} \}_{1:M}$. In this section, we propose using these draws to adaptively discover subgroups of patients with different cost-effectiveness profiles. This is ``adaptive'' in the sense that the number of clusters is not pre-specified, but grows or shrinks as the model adapts to the data complexity. Subgroup discovery is a policy-relevant endeavor since current CEA practice tends to focuses on marginal, population-level analyses - even if there is significant variation in the target population. Existing approaches to heterogeneity \cite{hahn2017, henderson2017, Athey2019} focus on computing ITEs and use post-hoc heuristic procedures to characterize this heterogeneity across pre-defined subgroups - rather than proposing subgroups adaptively.

\begin{figure} 
    \centering
    \begin{subfigure}[t]{0.32\textwidth}
        \centering
        \includegraphics[width=\linewidth]{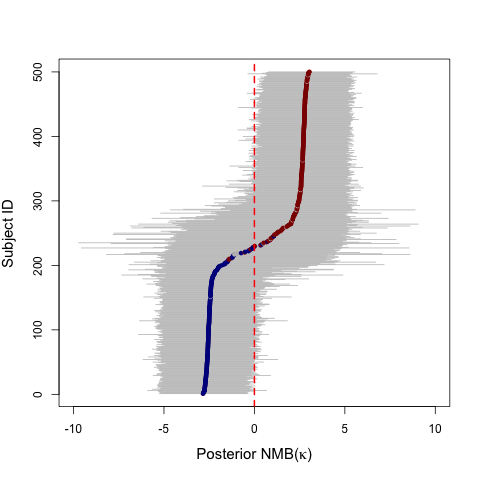} 
        \caption{} 
        \label{fig:indnmb}
    \end{subfigure}
    \hfill
    \begin{subfigure}[t]{0.32\textwidth}
        \centering
        \includegraphics[width=\linewidth]{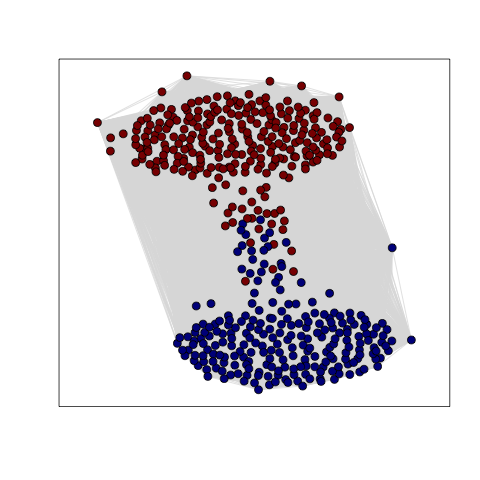} 
        \caption{} 
        \label{fig:clustgraph}
    \end{subfigure}
        \hfill
    \begin{subfigure}[t]{0.32\textwidth}
        \centering
        \includegraphics[width=\linewidth]{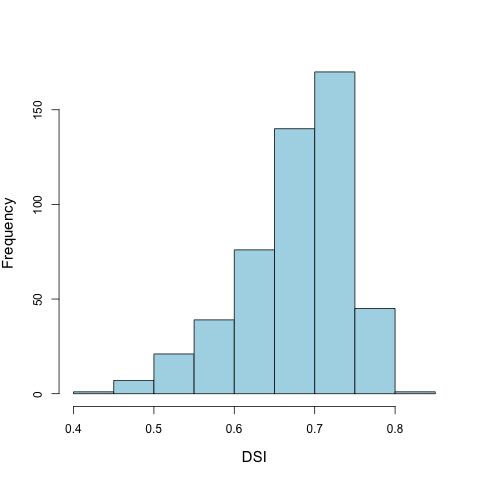} 
        \caption{} 
        \label{fig:dsipost}
    \end{subfigure}
    \vspace{.1in}
    \label{fig:asd}
    \caption{Clustering results from EDP-GP fit using synthetic data with two latent cost-effectiveness clusters. Here, EDP-induced clusters on the joint distribution capture differences in $NMB$. Panel \ref{fig:indnmb} shows posterior point and 95\% interval estimates $\Psi_i$ (with $\kappa=1$). Colors indicate posterior model cluster assignment, $c^*_{1:n}$. Panel \ref{fig:clustgraph} visualizes the posterior probability matrix $\mathcal{P}$. Panel \ref{fig:dsipost} is the posterior distribution of $DSI$ - indicating that about 70\% of the variation in subject-level $\Psi_i$ is explained by the EDP clustering. However, this need not be the case. The EDP clusters may be capturing complexities unrelated to NMB. While this is desirable to obtain a good fit to a complex distribution, it means the clusters have no substantive meaning. The $DSI$ is necessary to distinguish between these scenarios.}
\end{figure}

Using the given MCMC outputs for subgroup discovery is challenging for two reasons. First, the vector of cluster assignment labels, $c_{1:n}^{(m)}$, have no meaning across MCMC iterations - making it difficult to determine the posterior mode partition. This is known as label switching \citep{stephens2000}. To illustrate, consider that a new cost-effectiveness cluster forms in iteration $m+1$ and all subjects previously in another cluster are re-assigned to this new cluster. In this case, even though the assignment has changed, the underlying composition of the cluster did not. As a solution, we propose to keep track of the $n\times n$ adjacency matrix $\mathcal{C}^{(m)}$, where the $ij^{th}$ element, $\mathcal{C}_{ij}^{(m)}$, is a binary indicator of subject $i$ and $j$ being in the same cost-effectiveness cluster at iteration $m$. Note that this is just the vector $c_{1:n}^{(m)}$ re-arranged into a matrix. Taking the elementwise mean of this matrix across the $m$ posterior draws yields a probability matrix $\mathcal{P} = (1/M) \sum_m \mathcal{C}^{(m)}$ where $ij^{th}$ element, $\mathcal{P}_{ij}$, is the posterior probability of subject $i$ and $j$ being in the same cost-effectiveness cluster. To get a hard clustering assignment, we then search draws, $\{ c_{1:n}^{(m)} \}_{1:M}$,  for the assignment that is ``closest'' to $\mathcal{P}$. That is, we search for $c^*_{1:n} = \argmin_{m} ||\mathcal{C}^{(m)} - \mathcal{P}||$, where $||\cdot||$ is some matrix norm. As in earlier papers on Bayesian clustering, here we adopt ``Binder's Loss'' $|| \cdot || = \sum_{i,j} ( \mathcal{C}^{(m)}_{ij} - \mathcal{P}_{ik} )^2$ \citep{Binder1978, Dahl2006}. This essentially approximates the posterior mode of the EDP-induced partition, $\mathcal{P}$. Figure \ref{fig:clustgraph} visualizes $\mathcal{P}$ from an illustrative synthetic example as a weighted graph where each subject is a node and the length of vertices connecting two nodes are inversely proportional to $\mathcal{P}_{ij}$. Subjects with low posterior probability of being in the same cost-effectiveness cluster are far apart on the graph. Such figures are good tools for assessing uncertainty in posterior mode assignments, $c_i^*$. For instance, the points between the group of dark red and blue clusters represent subjects with highly uncertain mode assignments. The covariate effects of these subjects look just as similar to the well-separated dark blue points as they do to the well-separated dark red points.

A second challenge with using the assignments for subgroup discovery is that the EDP clusters are not explicitly designed to cluster on $NMB$. The clustering is driven by the complexity of the joint cost-effectiveness distribution. This is necessary for a flexible joint distribution estimate, but may not translate into meaningful $NMB$ clusters. For instance, consider a bimodal cost-effectiveness distribution with two groups having very different mean costs. However, the difference in costs between treatment groups in both clusters may be the same. In this case, the EDP will likely introduce two clusters with similar $NMBs$. This begs the question: are the clustering results detecting subgroups with different cost-effectiveness profiles? To answer this question, we propose a posterior Differential Subgroup Index ($DSI$) that, at each MCMC iteration, computes the proportion of the total variation in the ITEs, $\Psi_i^{(m)}$, that is explained by the cluster partition in that iteration. First, define the mean $NMB$ in subject $i$'s cluster at iteration $m$: $\bar \Psi_{i}^{(m)} = \frac{1}{\sum_j I(c_j^{(m)} = c_i^{(m)})} \sum_{j \mid c_j^{(m)} = c_i^{(m)}} \Psi_j^{(m)}$. Then the $DSI$ measure is, 
\begin{equation} \label{eq:dsi} 
	DSI^{(m)} = \frac{\sum_i \Big( \bar \Psi_{i}^{(m)} - \Psi^{(m)} \Big)^2 }{ \sum_i (\Psi_i^{(m)} - \Psi^{(m)})^2 } 
\end{equation}
This intuitively plays the same role as a regression $R^2$ statistic. Across the $m$ iterations, we have a set of draws for this statistic, $\{ DSI^{(m)} \}_{1:M}$, which reflects our uncertainty about how well the clustering is capturing heterogeneity in $NMB$. A posterior distribution for $DSI$ concentrated near 1 suggests that the EDP-induced clustering explains nearly all of the variation in the subject-specific $NMBs$. This implies that the EDP-induced clustering at the joint cost-effectiveness level is capturing variation at the NMB level. Figure \ref{fig:dsipost} plots the posterior distribution for $DSI$ for an illustrative synthetic example generated with two cost-effectiveness clusters. We can then summarize our data along the mode partition, $c_{1:n}^*$. For instance, in the synthetic example, we can create a table summarizing the observed costs, survival, and covariate distributions of the two identified clusters. These can be used to motivate future cost-effectiveness studies targeting these subgroups. The $DSI$ also provides context for our marginal posterior estimate, $\Psi$. A high $DSI$ indicates that a marginal estimate is not capturing substantial treatment effect heterogeneity detected by the EDP posterior.

\section{Assessing Frequentist Properties via Simulation}
In this section we report results of several simulation experiments exploring the frequentist properties (i.e. bias, coverage, and precision) of our posterior mean and interval estimates for $\Psi$ under a variety of settings. These results are reported in Table \ref{tab:simres}. We simulate data with one continuous confounder, four binary confounders, and a binary treatment. We simulate survival times conditional on treatment and confounders from a Weibull distribution. Survival times are censored by censoring times that also follow a covariate-dependent Weibull distribution. We simulate an outcome from a true $Y$ distribution of either a Gaussian or Log-Normal, with confounder- and treatment-dependent means. Data were simulated under low (5\%) and high (20\%) covariate-dependent censoring. For each of these, we simulate under a parametric and bimodal setting. Under the parametric setting, the joint distribution is unimodal - leading to a simple joint cost-survival distribution. Under the bimodal setting, we simulate data from a mixture of two cost-effectiveness distributions, each having different covariate effects in the cost and survival time models. In each of these eight settings, we simulate 200 datasets with 1500 subjects each. Details about the data generation are given in Appendix \ref{ap:simdetails}. 

We include the doubly-robust estimator (DR-SL) of \cite{Li2018} as a comparator. This approach involves estimating separate models for conditional mean cost and conditional mean survival time via super learner. Predictions from these models are weighted by the product of the inverse probability of treatment and inverse probability of censoring. We estimate the former using a correctly specified logistic regression - which suggests the DR estimate will be consistent but may still have substantial bias in finite samples if the models are inadequate. For the latter, we note that Li et al. did not consider the covariate-dependent censoring in their analysis. Instead, they estimate the probability of censoring in both treatment groups separately via Kaplan-Meier. Li et al. suggest using a discrete-time failure model in situations with covariate-dependent censoring. Here, we contribute to the literature by implementing this suggestion using a logistic regression. In the super learner libraries, we include regression trees, generalized additive, linear models, as well as elastic net generalized linear model (GLMnet). As recommended by Li et al., we using the bootstrap BCa interval for inference.

\input{simulation_results.tex}

For the EDP-GP, we run using independent Gaussian base distributions for $G_0$ that are null centered with flat priors, relative to the data variance. Importantly, we use a local conditional Gaussian model for $Y$. We set $\lambda_0^*$ to an exponential (constant) hazard. Additional details on DR-SL and EDP-GP settings are provided in Appendix \ref{ap:simdetails}. To summarize, the unimodal setting with Normally distribution $Y$ is the most favorable setting for our method since the Gaussian data generating model matches the local Gaussian model we specify. In principle, all of these settings are quite favorable to the DR-SL method since we correctly specify the propensity score model. The log-Normal setting is the least favorable to our method since our local Gaussian model is misspecified. Notice that in all censoring and $Y$ distribution settings, the parametric data generating process yields low bias and close to nominal coverage for both methods. This is as expected since both are highly flexible models, they should perform well in simple settings. Note however, that the models diverge in the more complicated, bimodal setting. In the bimodal log-Normal setting, the DR-SL exhibits higher bias with a larger interval width relative to EDP-GP. Similarly, in the bimodal Gaussian setting, the DR-SL model exhibits particularly high bias - 11\% and 16\% in the low and high settings, respectively. The main challenge with DR-SL is that the underlying super learner fails to capture biomodality in the cost-effectiveness joint distribution. In contrast, the EDP partitioning picks up the bimodality - modeling each mode with separate parameters to attain a better overall fit. Finally, note that EDP-GP intervals tend to be narrower across settings.

\section{Cost-efficacy of Endometrial Cancer Treatment}

We apply our BNP method to assess the cost-effectiveness of adjuvant chemotherapy (CT) versus radiation therapy (RT) for the treatment of endometrial cancer and compare our results to the DR-SL estimate. The target population of interest are women over the age of 65 who were diagnosed with endometrial cancer before undergoing hysterectomy. Within three months after hysterectomy, patients are assigned to either adjuvant RT or CT. We select a cohort of women over the age of 65 who were diagnosed with endometrial cancer between 2000 and 2014 in the SEER-Medicare database. The first treatment after three months of diagnosis was recorded. A maximum of $\tau=24$ months of follow-up after hysterectomy was available in this data cut. Total costs accrued by Medicare (including inpatient, outpatient, hospice, and pharmaceutical costs) were recorded along with their survival/censoring status. Covariates which are known drivers of treatment assignment (age, comorbidities, cancer stage) were extracted. Table \ref{tbl:bsl} displays summary statistics for the sample. Notably, the 2-year survival is slightly lower in the CT arm (93\% vs. 94.5\%), and average total costs higher in the CT arm (51.3 vs. 42.6). This suggests worse cost-effectiveness for CT relative to RT. However, there is significant uncertainty associated with these numbers that should be quantified. Moreover, the cohorts differ substantially in terms of observed characteristics at treatment assignment. For instance, the radiation arm has a greater proportion of patients with baseline International Federation of Gynecology and Obstetrics (FIGO) stage of IB - which is more severe than IA and I-NOS. Similarly, RT harm has fewer comorbidities - with 57\% (vs. 54\%) having Charlson Comorbidity Index of zero. These differences could differentially affect adjuvant therapy assignment and cost-efficacy. 

\input{bsl_table.tex}

We use our EDP-GP approach to compute posterior point and interval estimates for NMB while adjusting for differences in observed covariates. We specify the local cost distribution, $p(Y_i \mid T_i, \delta_i, X_i, \omega_i )$, to be a log-normal distribution with parameters $\omega_i = (\beta_i, \phi_i)$. The local regression is
$$ E[Y_i \mid T_i, \delta_i, A_i, L_i, \omega_i] = \exp\big\{ (1, L_i, A_i, T_i, \delta_i)' \beta_i + \phi_i/2\big\} $$ 
This local log-normal distribution respects the non-negative nature of costs, while allowing us to capture skewness. In the model, $L_i$ includes household income, Charlson index, and FIGO. FIGO is included as a categorical covariates, while the others are treated as continuous. We let $A=1$ indicate assignment to chemotherapy with radiation being reference.

We set prior $G_0$ as discussed in Section \ref{sc:hyperparms}: $G_{0\omega}(\beta_i, \phi_i) = N(\hat{\beta}, \hat\Sigma) IG(a_0, \hat\phi(a_0 -1) )$. Here, $\hat{\beta}$ are OLS estimates using $\log(Y)$ as the outcome and $\hat \Sigma = diag(1, .01^2, \dots, .01^2)$. Note the latter appears overly informative, but is actually fairly wide on the exponentiated scale. That is, a prior variance of $1$ implies that mean costs as large as $exp(1\cdot 1.96)\approx 7$ times the empirical mean cost are plausible. Similarly, a prior variance of $.01$ implies covariate effects of as large as $2\%=1-exp(1.96*.01)$ are \textit{a priori} plausible in the absence of data. For $\phi$, note that the variance of the log-Normal random variable, $Z$, is $Var[Z] = (e^\phi - 1) E[Z]^2$, which implies $\phi = \log[ Var[Z]/E[Z]^2 + 1]$. This motivates setting $\hat \phi = \log[ \hat s^2 /{\bar y}^2 + 1] $, where $\hat s^2$ and $\bar y$ are the marginal variance and mean of the observed cost values. We set $a_0=1000$, which anchors the prior around the empirical estimate. For the effectiveness model we again follow Section \ref{sc:hyperparms} and set $G_{0\theta}(\theta_i) = N( \hat\theta_{PH}, I)$. We center the GP priors around an empirical estimate $\lambda_0^*(t) = \hat \lambda \approx .001$ with $b=2000$ and $\xi=4000$. Here, $\xi$ is on the order of the sample size - signifying strong AR(1) smoothing. The value $b$ is about half of $\xi$ - putting equal $\textit{a priori}$ weight on the prior hazard $\lambda_0^*(t)$ and the previous hazard at time $t-1$.

We run three MCMC chains in parallel for 5,000 iterations and discard the first 3,000 draws  of each chain as burn-in. We initialize each chain with different numbers of initial cost and effectiveness clusters and check that the chains converge to each other regardless of this initialization. This yields a total of 6,000 draws which we use for posterior inference. Other details and assessments of convergence are provided in Appendix \ref{ap:datadetails}.

\begin{figure} 
    \centering
    \begin{subfigure}[t]{0.32\textwidth}
        \centering
        \includegraphics[width=\linewidth]{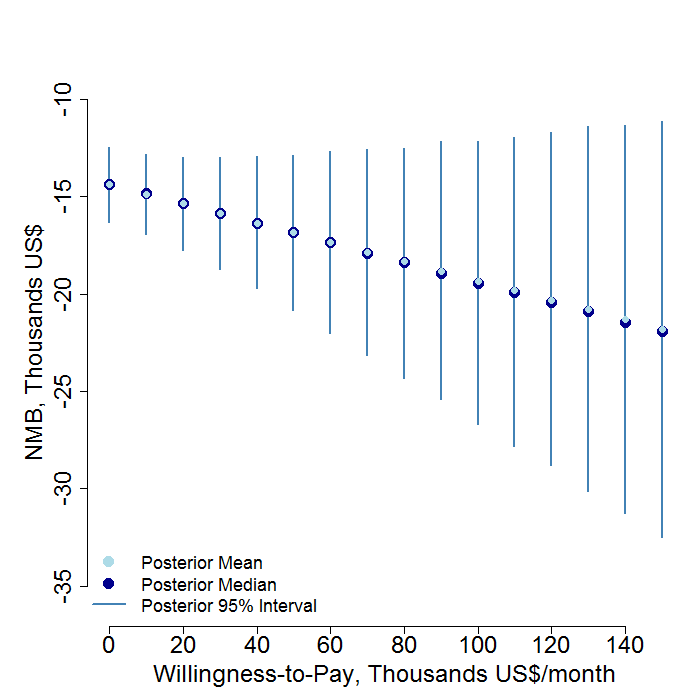} 
        \caption{} 
        \label{fig:nmbest}
    \end{subfigure}
    \hfill
    \begin{subfigure}[t]{0.32\textwidth}
        \centering
        \includegraphics[width=\linewidth]{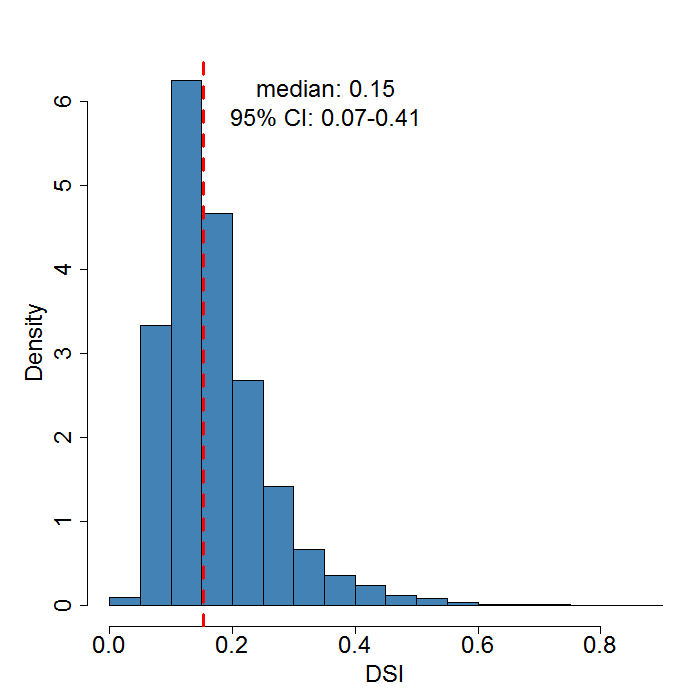} 
        \caption{} 
        \label{fig:dsiest}
    \end{subfigure}
        \hfill
    \begin{subfigure}[t]{0.32\textwidth}
        \centering
        \includegraphics[width=\linewidth]{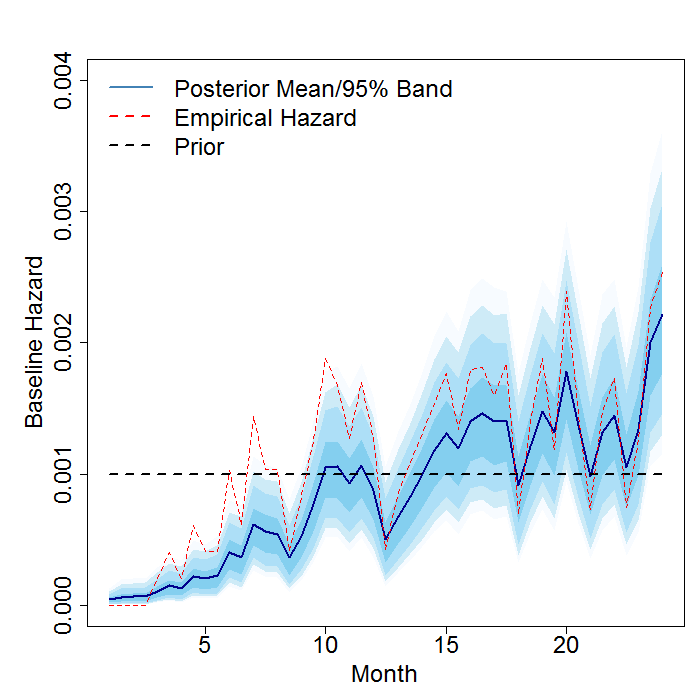} 
        \caption{} 
        \label{fig:hazplot}
    \end{subfigure}
    \vspace{.1in}
    \label{fig:haz}
    \caption{Posterior estimates of (a) NMB for various willingness-to-pay for each additional \textit{month} of survival, $\kappa$. The posterior distribution of $DSI$ in (b) shows that about 15\% of the variation in the individual-level NMBs is explained by the EDP induced clustering. This suggests the treatment effect may be relatively homogeneous and the NMB is a good overall average effect measure. In panel (c) we have plotted the posterior baseline hazard curve along with 95\%, 90\%, and 80\% credible bands in successively darker shades. Notice that posterior estimate is smoother version of the empirical estimate hazard in red. It is a posterior compromise between the empirical hazard and the prior constant hazard. }
\end{figure}

We estimate a 2-year NMB of chemotherapy over radiation to be $-\$14.5$ thousand, with $95\%$ CI $[-\$16.6,-\$12.7]$. This assumes a willingness-to-pay of about $\kappa= \$4167$/month, or $\$50,000$/year of life gained - which is standard in cost-effectiveness analyses. This is roughly consistent with the unadjusted comparison in Table \ref{tbl:bsl}, where average total costs among chemotherapy patients was higher by about $\$9,000$. Figure \ref{fig:nmbest} shows average NMB as a function of $\kappa$ for various $\kappa$ values. Recall that by definition NMB is a linear function of $\kappa$. The intercept at $\kappa=0$ shows an NMB that captures differences in cost only (efficacy has zero value). The negative y-intercept here reflects that even if we do not value efficacy, chemotherapy is more expensive than radiation after covariate adjustment. The negative slope of the curve reflects that adjusted efficacy (i.e. survival benefit) of chemotherapy is lower. However, the slope here is quite small, suggesting a very small difference in efficacy. This is consistent with unadjusted results - recall from Table \ref{tbl:bsl} that 2-year survival is slightly lower among chemotherapy patients.

In terms of clustering, we compute $c^*_{1:n}$ as given in Section \ref{sc:subgroup} and find that about 86\% of the observations are grouped into two posterior mode clusters. However, in Figure \ref{fig:dsiest} we see that only about $15\%$ of the variation in the individual-level NMBs is explained by the EDP-induced partition - which suggests these clusters are not very meaningful for cost-effectiveness. This indicates low posterior evidence of treatment effect heterogeneity, suggesting average NMB may fairly characterize the cost-effectiveness profile. Finally, Figure \ref{fig:hazplot} shows the posterior estimate of the baseline hazard. Since continuous covariates were normalized, this represents the hazard among patients with average household income and age with Charlson index of zero and FIGO II/II-NOS. This has no explicit causal interpretation but is illustrative of the Gamma process. Notice our posterior has moved away from the constant hazard prior and towards the empirical (Nelson-Aalen) estimate shown in red. The informative AR(1) shrinkage results in a smoother posterior curve that penalizes large swings in the empirical hazard.

For comparison, we also ran the DR-SL approach where propensity score model, mean survival time model, and mean cost model were all estimated using super learner. Regression trees, GLMnet and GLM were included as candidate learners and 95\% BCa intervals were estimated using 5,000 bootstrap iterations. For willingness-to-pay $\kappa= \$4167$/month, DR-SL estimates a 2-year average NMB of $-\$11.8$ with 95\% CI $[ -\$19.1, -\$6.0 ]$ in thousands. This is similar to our estimate of $-\$14.5 \ [-\$16.6,-\$12.7]$, but the DR interval is wider. More details on the DR-SL implementation are given in Appendix \ref{ap:datadetails}, including a full plot of average NMB from DR-SL as in Figure \ref{fig:nmbest}. For even large willingness-to-pay values of up to 300 thousand USD per year, both approaches find a negative NMB with intervals excluding zero. This supports the relative cost-effectiveness of radiation over chemotherapy adjuvant therapy over two years.

\section{Discussion}

Cost-effectiveness is statistically challenging due to the complexities of the joint distribution of cost and survival time, such as skewness, censoring, and multi-modalities. Moreover, estimation of policy-relevant estimands with causal interpretation is complicated by confounding in observational studies. Robust causal inference for cost-effectiveness requires flexible modeling that accounts for these complexities while adjusting for confounders. In this paper, we outlined a nonparametric Bayesian solution that leverages the Gamma and enriched Dirichlet process priors to model the joint distribution of cost and survival time. We proposed cost-effectiveness estimands with causal meaning and identified them under suitable causal assumptions. We showed how our model can be used in a Bayesian g-computation procedure that draws from the posterior of the causal effect. Finally, we show that the partition induced by the EDP can be used to explore cost-effectiveness heterogeneity and introduced the $DSI$ diagnostic statistic for assessing how well this partition captures heterogeneity.

In simulations, we demonstrated that our procedure has adequate frequentist properties (bias, coverage, etc.) in a variety of scenarios. In complex settings, it can be comparable and, at times, outperform existing doubly-robust methods. Across almost all settings, the EDP-GP produces NMB estimates with narrower interval widths relative to the DR-SL estimates. In the data analysis, the DR-SL approach also yields wider intervals. One driver of this is the relative inefficiency of the DR-SL approach. This method only uses data from patients who are not censored and weights their contributions by the inverse probability of being uncensored. In contrast, our method uses censored patients, since they still inform the total cost distribution at their observed time. Another feature with the DR-SL is that it is a weight-based estimator (weighted both by inverse probability of treatment and censoring), which are known to be quite variable if the probability of treatment are near the bounds within subgroups. Since the EDP-GP approach is model-based, it provides more smoothing under these conditions. Finally, the bootstrap inference procedure used in the DR-SL approach can be difficult to implement in practice, where sparsity among categorical covariates leads to the occasional pathological bootstrap resample (e.g. with rank deficient matrix). This is in contrast to full posterior inference via the Bayesian bootstrap which can be more stable.

Finally, we see at least two avenues of future work and extensions. First, in our paper, we consider a setting with a single baseline treatment. This allows us to estimate the cost-effectiveness of baseline treatments, which are highly relevant in many settings. However, we may also wish to estimate the cost-effectiveness of time-varying treatment \textit{regimes}, in addition to the effect of the initial baseline treatment. Flexible causal estimation in these settings is more complex and should be explored. Second, there has been much work on improving the computational scalability of posterior inference on Dirichlet process models, including both approximate inference via Variational Bayes and parallel MCMC procedures. Future work developing scalable inferential procedures for joint-modeling with EDPs can be useful.

%
%

\section*{Acknowledgements}

Dr Emily Ko was partially supported by Grant 124268-IRG-78-002-35-IRG from the American Cancer Society, the George and Emily McMichael Harrison Fund, Penn Presbyterian Harrison Fund of the University of Pennsylvania Hospital Obstetrics and Gynecology Department. The analysis used the linked SEER-Medicare database and we acknowledge the efforts of the Applied Research Program; National Cancer Institute; Office of Research, Development and Information; Centers for Medicare and Medicaid Services; Information Management Services; and SEER program tumor registries in the creation of the SEER-Medicare database.
 


\bibliographystyle{imsart-nameyear} 
\bibliography{BNPCE} 

\newpage
\begin{appendix}
\section{Identification of Causal Net Monetary Benefit} \label{ap:identification}

Recall that we are interested in estimating $\Psi = E[MV^{1,0} ] - E[MV^{0,0}]$, where the expectation implicitly conditional on the parameters governing the joint cost-survival distribution. We can identify each term of $\Psi$. Starting with an iterated expectation over $L$, 
\begin{align*} 
	E[MV^{a,0}] & = E_\mathcal{L}[ E_{\mathcal{Y}, \mathcal{D} } [MV^{a,0} \mid L, \omega_{1:n}, \theta_{1:n}, \lambda_0 ]] \\
		      & = E_\mathcal{L}[ E_{\mathcal{Y}, \mathcal{D} } [MV^{a,0} \mid, A=a, \delta=0, L, \omega_{1:n}, \theta_{1:n}, \lambda_0 ]] \\
		      & = E_\mathcal{L}[ E_{\mathcal{Y}, \mathcal{D} } [MV \mid A=a,  \delta=0, L, \omega_{1:n}, \theta_{1:n}, \lambda_0 ]] \\
		      & = \int_\mathcal{L} E_{\mathcal{Y}, \mathcal{D} } [MV \mid A=a,  \delta=0, L, \omega_{1:n}, \theta_{1:n}, \lambda_0 ] dP(L) \\
		      & = \int_\mathcal{L} \int_{\mathcal{Y}, \mathcal{D} } (D\kappa - Y) p(Y, T \mid A=a, L, \delta=0, \omega_{1:n}, \theta_{1:n}, \lambda_0) dP(L) \\
\end{align*}
Note above, $\mathcal{Y}$ and $\mathcal{D}$ are the spaces we integrate over. This last line is each term of Equation \eqref{eq:mcmv}. The second line follows from joint ignorability (IA.1), allowing us to condition on $A=a, \delta=0$ after first conditioning on $L$. The third line follows from joint consistency, IA.2, allowing us to drop the superscripts on monetary value. These are extensions of the usual conditional ignorability and consistency assumptions under censoring \citep{robins2000} extended to handle a bivariate cost-survival time outcome. The interference assumption, IA.4, allows us to write $MV^{a_{1:n},0_{1:n}}=MV^{a,0}$. That is, each subject's potential monetary value is independent of others' treatments or censoring status. Said another we, we learn nothing about someone else's potential monetary value by learning another's treatment assignment. Joint positivity, IA.3, is requires so that we do not condition on a zero-probability event in the second equality. The expression above identifies a causal estimand that is purely a function of unknown parameters. Thus a posterior distribution over the parameters induces a posterior distribution over monetary value.

\section{Posterior Computation} \label{ap:mcmc}
\subsection*{Gamma Process Prior Specification}
This appendix provides additional details for updating the baseline hazard model with a dependent Gamma process prior \citep{barajas2002}. Much of this is a detailed overview of the results established by Nieto-Barajas and others in their 2002 paper and outlined in documentation of the \textbf{BGPhazard} R package. We provide an abbreviated presentation adapted to the context of our joint model for the reader's convenience.

Consider observing right-censored survival time data for $i=1,\dots, n$ subjects with survival time $T_i$ and death indicator $\delta_i$. Consider a partition, $\{\tau_v\}_{v=1:V} $, of the time interval such that $0 < \tau_1 < \tau_2 < \dots < \tau_V $ where $\tau_V > max_i(T_i)$. In a setting with fixed study end, $\tau$, we could set $\tau_V = \tau$. In this case we consider equally-spaced interval such that $\Delta_v = \tau_v - \tau_{v-1}$ for all $v$. A piecewise constant hazard model can be defined as
$$\lambda_0(t) = \sum_{v=1}^{V} \lambda_{0v}I(\tau_{v-1} < t \leq \tau_v)$$
If \textit{a priori} the baseline hazard $\lambda_0(t) \sim GP(b\lambda_0^*, b, \xi=0)$, then the hazard rate in each interval follows $\lambda_{0v} \sim Gam( b\lambda_{0v}^*, b )$, where the first argument in the shape and the second argument is the rate. In the shape, we've defined $\lambda_{0v}^* = \{\Lambda_0^*(\tau_v) - \Lambda_0^*(\tau_{v-1})\}/\Delta_v$, where $\Lambda_0^*$ denotes the prior cumulative hazard. Thus the prior mean hazard at each interval is $E[\lambda_{0v}] = \lambda_{0v}^* $. This is known as the independent Gamma process prior because the hazard at each increment is independent \textit{a priori}. The \textit{dependent} Gamma process of Nieto-Barajas extends this process to introduce dependence between hazards in nearby increments - providing a smoother estimate that is less dependent on choice of time partition. They do this by introducing latent processes $\{c_v\}_{1:V}$ and $\{u_v \}_{1:V}$ and is denoted with GP, as above, but with $\xi>0$. The process is initialized with $\lambda_1 \sim Gam( b\lambda_{01}^*, b )$. Now for $v\in\{1,2,\dots \tau-1\}$, we have $u_v \mid \lambda_v, c_v \sim Pois(c_v \lambda_{0v} )$ and $\lambda_{0v+1} \mid u_v, c_v \sim Gam( b\lambda_{0v+1}^* + u_v, b + c_v )$. The conditional prior mean of this process is $$E[\lambda_{0v} \mid \lambda_{0v-1} ] = \frac{b \lambda_{0v}^* + c_{v-1}\lambda_{0v-1}^* }{ b + c_{v-1} } $$
So the prior mean baseline hazard rate in current interval $v$ is a weighted average of the prior baseline hazard rate, $\lambda_{0v}^*$, in the current time interval and the prior baseline hazard rate in the previous time interval, $\lambda_{0v-1}^*$. This is the induced AR(1) smoothness of the dependent Gamma Process. Following, Nieto-Barajas we place a hyperprior on $\{c_v\}_{1:V}$, assuming $c_v  \mid \xi \stackrel{iid}{\sim} Exp(\xi)$. Where the prior mean is $E[c_v] = \xi$. The magnitude of $\xi$ (relative to b) controls the aggressiveness of the prior AR(1) shrinkage. if $\xi >> b$, then on average $c_{v-1}>>b$ at all intervals $v$, meaning that $E[\lambda_{0v} \mid \lambda_{0v-1} ] \approx \lambda_{0v-1}^*$. Similarly, if $\xi << b$, then $E[\lambda_{0v} \mid \lambda_{0v-1} ] \approx \lambda_{0v}$ - i.e. almost no shrinkage to the previous hazard. It can be shown above that setting $\xi=0$ above reduces this to the independent Gamma process.

Thus, the notation $\lambda_0 \sim GP(b\lambda_0^*, b, \xi)$ denotes this prior for the piecewise constant model $\lambda_0(t)$. Specifically, the joint prior is
\begin{equation}  \label{eq:gppr} \small
p(\lambda_{01:V}, c_{1:V}, u_{1:V} \mid b, \xi) = p(\lambda_1)p(u_1 \mid \lambda_{01}, c_1) \prod_{v=2}^V p( u_v \mid \lambda_v, c_v) p( \lambda_{0v} \mid u_{v-1}, c_{v-1}) \prod_{v=1}^V p(c_v \mid \xi)
\end{equation}
With hyperparameters $b$, $\xi$, and $\lambda_0^*$. Notational dependence on $\lambda_0^*$ has been suppressed for compactness. This can be combined with the likelihood for the observed data to obtain conditional posteriors for each of the three parameter blocks, $\lambda_{01:V}, c_{1:V}$, and $u_{1:V}$. We discuss likelihood construction in the next section.
b
\subsection*{Gamma Process Likelihood Construction}
Now we consider the $GP(b\lambda_0^*, b, \xi)$ prior for the baseline hazard in a proportional hazard model $\lambda(t \mid X_i, \theta_i) = \lambda_0(t) \exp \big( X_i' \theta_i \big)$, where $\lambda_0(t) = \sum_{v=1}^{V} \lambda_{0v}I(\tau_{v-1} < t \leq \tau_v)$ . Specifically, our goal is to find the posterior $p( \{\lambda_{0v}\}_{1:V},  \{c_{v}\}_{1:V},  \{u_{v}\}_{1:V} \mid \mathcal{D} )$, where $\mathcal{D}$ indicates the observed data.

For convenience in presentation, define $\eta_i = X_i' \theta_i$. Also note that under the piece-wise constant model, the cumulative hazard is $ \Lambda_i(t) = \int_0^t \lambda_{0}(s) e^{\eta_i} ds = \sum_{v=1}^{V} \lambda_{0v} e^{\eta_i} \Delta_v(t)$. Here, $\Delta_v(t) = (t-\tau_{v-1})I(t\in (\tau_{v-1}, \tau_v]) + \Delta_v I( t>\tau_v)$.

 Conditional on $\theta_i$, standard survival likelihood construction with right-censored data yields
\begin{equation*}
	\begin{split}
		p(T_i \mid X_i, \theta_i, \delta_i, \lambda_{01:V}) =  \prod_{i \mid \delta_i=1}f(T_i \mid X_i, \theta_i)  \prod_{i \mid \delta_i=1} S(T_i \mid X_i, \theta_i)
	\end{split}
\end{equation*}
Subjects with an event contribute to the likelihood via the density, $f$, and censored subjects contributed via the survival function $S$, both of which can be expressed in terms of the hazard. Denote $\lambda_{0v_i}$ as the hazard rate of the increment in which subject $i$ died. The density evaluated at subject $i$'s death time is,
\begin{equation} \label{eq:tdens}
f(T_i \mid X_i, \eta_i)  = \lambda_0(T_i) e^{ - \Lambda_i(T_i) } = \lambda_{v_i} e^{\eta_i} \exp\Big\{ - \sum_{v=1}^{V} \lambda_{0v} e^{\eta_i} \Delta_v(T_i)  \Big\}
\end{equation} 
The survival function in terms fo the hazard is, 
$$ S(T_i \mid X_i, \theta_i ) = \exp\Big\{ - \Lambda_i(T_i) \Big\} = \exp\Big\{ - \sum_{v=1}^{V} \lambda_{0v} e^{\eta_i} \Delta_v(T_i)    \Big\}  $$
So the full likelihood is
\begin{equation}
\begin{split} \label{eq:gplik}
	p(T_i \mid X_i, \theta_i, \delta_i, \lambda_{01:V}) = \Big(\prod_{i\mid \delta_i = 1} \lambda_{0v_i} \Big) \exp \Big\{ \sum_{i\mid \delta_i=1} \eta_i \Big\} \exp\Big\{ - \sum_{v=1}^V \lambda_{0v} \Big(  \sum_{i=1}^n e^{\eta_i} \Delta_v(T_i ) \Big) \Big \}
\end{split}
\end{equation}

\subsection*{Gamma Process Posterior Updates}
The likelihood \eqref{eq:gplik} can be combined with the joint prior \eqref{eq:gppr} to obtain the following conditional posteriors distributions for $u_{1:V}$, $c_{1:V}$, and $\lambda_{01:V}$. Note all of these distributions are also conditional on data, $\mathcal{D}$. First, the conditional posterior distribution of $\{c_v\}_{1:V}$ is 
\begin{equation}
	  p(c_v \mid u_v, \lambda_{ov+1}, \lambda_{0v}) \propto
    \begin{cases}
      c_v^{u_v} \exp \Big \{ - (\lambda_{0v} + \lambda_{0v+1} + \frac{1}{\xi} ) c_v \Big \} (b + c_v)^{ \lambda_{0v+1}^* + u_v } &  v=1,\dots, V-1\\
      Gam(u_v + 1, \lambda_{0v} + \frac{1}{\xi}) &  v=V\\
    \end{cases} 
\end{equation}
For $v=1,\dots, V-1$ this update is not conjugate. We sample each $c_v$ separately using Adaptive Metropolis-Hastings with separate proposal variances for each $c_v$. The proposal variances are tuned every few iterations in the burn-in period to target a $23.4\%$ acceptance rate, which has been shown to be optimal in around 10-dimensional sampling problems \citep{roberts2001}. The latent process $\{ u_v\}_{1:V}$ can be updated from the following conditional posterior, 
\begin{equation}
	  p(u_v \mid c_v, \lambda_{0v+1}, \lambda_{0v}) \propto
    \begin{cases}
        \frac{ \big[ c_v \lambda_{0v}\lambda_{0v+1} (b + c_v) \big]^{u_v} }{\Gamma(u_v + 1)\Gamma( \lambda^*_{0v+1} + u_v )} &  v=1,\dots, V-1\\
       Pois(c_v \lambda_{0v}) &  v=V\\
    \end{cases} 
\end{equation}
Note here $u_v$ is integer-valued and non-conjugate for $v=1,\dots,V-1$. To sample from these conditional posteriors, we use grid sampling with a large grid of points $\{ 0,\dots, 10000 \}$. Finally, the conditional posteriors of the hazard rate in each interval is given by
\begin{equation}
	  p(\lambda_{0v} \mid - , D) =
    \begin{cases}
         Gam\big( d_1 + u_1 + \lambda_{01}^*, c_1 + b + \sum_{i=1}^n e^{\eta_i} \Delta_1(T_i) \big) &  v=1\\
         Gam\big( d_v + u_v + u_{v-1} + \lambda_{0v}^*, b + c_{v} + c_{v-1} +   \sum_{i=1}^n e^{\eta_i} \Delta_v(T_i)  \big) &  v=2, \dots, V\\
    \end{cases} 
\end{equation}
Above, $d_v$ is the number of deaths in interval $v$. Note that the conditional distribution is fully conjugate for all $v$ and can be sampled directly. Note also that this update is the only Gamma Process update that involves data. The processes $u_{1:V}$ and $c_{1:V}$ are latent and the updates do not involve data - but they do induce a dependence between the $\lambda_0v$, which now must be updated sequentially and in order. 

\subsection*{Concentration Parameters}
The two concentration parameters of the EDP, $\alpha_\theta$ and $\alpha_\omega$, are given $Gam(1,1)$ priors. We follow the implementation in \cite{Roy2018}. Details can be found in the supplement to their 2018 paper.

\subsection*{Monte Carlo Integration for Monetary Value}
\label{ap:mcinteg}

The expectation can be expressed as 
\begin{align*}
	\scriptsize
	\mu(a, 0) & = \kappa E[D \mid - ]  - \int_0^\tau \int_0^\infty  E[Y \mid D, - ]p(D \mid - ) dY dD \\
\end{align*}
Note we use ``$-$" to denote the conditioning set, which was made explicit in the main body of the paper.
\begin{itemize}
    \item The first term, $E[D \mid - ]$, (average death time within 2-years under treatment $a$) can be computed in closed form. Since we partition time interval (see Appendix \ref{ap:mcmc}) into $K$ intervals, the probability of dying in  interval $k$ is $ p(t \in [\tau_k, \tau_k+1] \mid - )$. Within each interval, death time is uniform - so mean is $\frac{\tau_{k+1} + \tau_k}{2}$.
    $$ E[D \mid - ] \approx \sum_{k=1}^K \frac{\tau_{k+1} + \tau_k}{2} \cdot p(t \in [\tau_k, \tau_k+1] \mid - )  $$
    At every iteration, $p(t \in [\tau_k, \tau_k+1] \mid - )$ is given by substituting the parameter draws in this iteration into Equation \eqref{eq:tdens}.
	\item Second term: For each subject, draw death interval proportional to $p(t \in [\tau_k, \tau_k+1] \mid - )$. Then, within each interval draw a death time $t^*$ uniformly within that interval. Compute $E[Y \mid T=t^*, - ]$ using this drawn value and the parameter draws in the current iteration.
\end{itemize}

\section{Simulation Details} \label{ap:simdetails}
\subsection*{Data Generation}
In the log-normal setting, we simulate data as follows. For subject $i=1,\dots, N$,
\begin{itemize}
	\item Simulate latent cluster membership: $c_i \sim Ber(p_c)$, a 5-dimensional confounder $L_i$. This vector contains one continuous confounder drawn from a standard Normal distribution in the first entry and four binary confounders draw from Bernoulli distribution with probability .5. 
	\item Simulate treatment: 
	$$A_i \sim Ber( expit(0 + (.1, .5,-.5, .5, -5)' L_i ) )$$
	\item Simulate survival time, $T_i$: from a Weibull distribution (using the proportional hazard parameterization) with shape 10 and scale $exp(\eta_i)$. Where
	$$ \eta_i = c_i\cdot[ (0,.1, -.1, .1, -.1)'L_i]  + (1-c_i)\cdot[ (1, -.1, .1,-.1, .1) ]  )  + (-3 + 2c_i) A_i $$
	Notice that the treatment effect on survival is bimodal, along with the covariate effects.
	\item Simulate a covariate-dependent censoring time: $C_i$, from the same Weibull as above.
	\item Simulate Observed time observed time: Draw $Z_i \sim Unif(0,1)$ and simulate censoring indicator  $\bar \delta_i = I(C_i < D_i)\cdot I(Z_i < p_\delta)$.  If $\bar\delta_i=1$, then $T_i = min(C_i, D_i)$.
	\item Simulate accumulated cost up to $T_i$: 
	$$Y_i \sim \log N\Big( mean = \mu_i, \ sd = .05 \Big)$$
	where 
	$$ \mu_i =  2 c_i +  (.1, .2, .2, .2,.2)'L_i - 2T_i + .3 A_i $$
	Here we have a bimodal cost distribution (different means depending on $c_i$) but homogeneous treatment effect on costs.
	\item Output observed data $D_i = (Y_i, T_i, \delta_i = 1 - \bar \delta_i, L_i, A_i)$.
\end{itemize}

In the Normal setting, we simulate data as above with the following modifications:
\begin{itemize}
	\item Simulate survival and censoring times time with $\log$ scale parameter
	$$ \eta_i = c_i\cdot[ (-1, .1, -.1, .1, -.1)'L_i]  + (1-c_i)\cdot[ (1, -.1, .1,-.1, .1) ]  )  + 2c_i \cdot A_i $$
	Note again that treatment and covaraite effects are bimodal (dependent on $c_i$).
	\item Simulate outcome data from a \textit{Normal} distribution with standard deviation .5 and mean 
	$$ \mu_i = 5 + 5c_i + (.1, .5, .5, .5, .5)'L_i  - 3A_i + T_i  $$
	\item Here the treatment and covariate effects on $Y$ are homogeneous.
\end{itemize}

We simulate each dataset with $N=1500$. In the bimodal setting, $p_c = .5$. In the parametric setting, the $p_c=0$ - so all subjects are from the same cluster. We set $p_\delta = .4$ in the high setting to target 20\% censoring and $p_\delta = .1$ in the low setting to target 5\% censoring. For each setting Normal/log-Normal -$p_\delta$-$p_c$ combination, we simulate 200 such datasets. 

\subsection*{EDP-GP Prior Settings}
First we discuss the settings for the log-Normal data generating mechanism. For the Gamma Process prior, we partition the interval from $[0, max(T_i)]$ into equal size increments of $.1$. We set $\xi=1e-6$ to be quite small (very flat) to allow the likelihood to drive the posterior estimate. We set $b=\xi$ thus inducing an AR1 dependence between increments that is as informative as the shrinkage towards $\lambda_0^*$, which we set to an exponential hazard with rate 400 - close to the average empirical hazard rate across time points. Notice the actual baseline hazard is generated from a Weibull, so our prior is deliberately misspecified as it likely would be in practice. 

The prior on $\theta_i$, $G_{0\theta}$ is set to a multivariate Gaussian with zero mean vector and diagonal covariance $3^2 I_6$. Where $I_6$ is the $6\times 6$ identity matrix, where $6$ is the number of covariates (5 confounders and one treatment indicator). This is flat on the hazard ratio scale.  

Since we fit a Gaussian conditional model for $Y$, the prior $G_{0\omega}$ is a product of a prior on the covariate effects and prior on the variance. Regarding the former, we again use a multivariate Gaussian with zero mean vector and covariance $3^2 I_7$, where the identity matrix has a diagonal entry for the five confounders, treatment indicator, and observed time. This is fairly flat relative to the true conditional outcome variance (on log scale) of $.05^2$. The prior for the variance is set to an inverse gamma distribution. In the bi-modal setting we set this distribution to have shape and scale equal to 20. This centers the prior variance around 1. In the parametric/unimodal setting we use a slightly tighter prior around 1 - with shape and rate equal to 100. These tighter settings like 20 and 100 help regularize the Gaussian model we fit to the skewed $Y$ data.

For the Normal data generating mechanism much of the settings above is the same. We only change the shape parameter of the inverse gamma distribution on the conditional cost variance to be 5 with a rate of 20. This is a fairly flat prior.

For each data set, we run the MCMC sampler for 7000 iterations and discard the first 2000 as burn-in. This yields 5,000 posterior draws which we use for inference about NMB. In all settings, we initialize the model with three $\omega$ clusters, each having three $\theta$ sub-clusters. This initialization is very different from the true data generating mechanism that either generates data from a single $\omega-\theta$ cluster and two $\omega$ (top-level) clusters. 

Since we fit a Gaussian model, each cluster's conditional $\omega$ posterior is conjugate with our Normal-Inverse-Gamma prior. This is a simple update. For the $\theta$ cluster parameters we use a Metropolis update with Gaussian jumping distribution. The jumping covariance is identity with .1 along the diagonals. Similarly, we use a Metropolis step to update $\{ c_{1:v} \}_{1:V}$ (see Appendix \ref{ap:mcmc}) at each step. Each $c_v$ is updated from an independent Gaussian jumping distribution with variance .5. We adapt both of these jumping distribution variances every 25 iterations starting from iteration 50 and ending at iteration 200 to target an acceptance rate of 23.4\% per \cite{roberts2001}.

\subsection*{Doubly-Robust Implementation}
Here we describe the doubly-robust NMB estimator of \cite{Li2018} implemented in our simulations. The cost and survival time models are estimated using super learner with regression trees, generalized additive models, generalized linear models, and GLM-Net included in the ensemble. We use a correctly specified logistic regression for the treatment model. This is quite generous since doubly-robust estimators are guaranteed to be consistent with a correctly specified treatment model (though the convergence rate can be quite slow if the outcome model is very misspecified.). 

Since we have covariate dependent censoring, we estimate the inverse censoring weights using a discrete-time failure model as described in Section 3.1.1 of their paper. To summarize, these weights are computed using estimates of the probability of censoring at each time point, conditional on not having been censored before that time point. This is estimated using a logistic regression of a censoring indicator at each time point on simulated confounders, treatment and time-level fixed effects. Intervals are computed using a 95\% BCa interval after 1502 bootstrap iterations (BCa intervals require more bootstrap iterations than observations in the sample).

\section{Data Analysis Details} \label{ap:datadetails}

\begin{figure}
	\centering
	\begin{subfigure}[t]{0.48\textwidth}
		\centering
		\includegraphics[width=\linewidth]{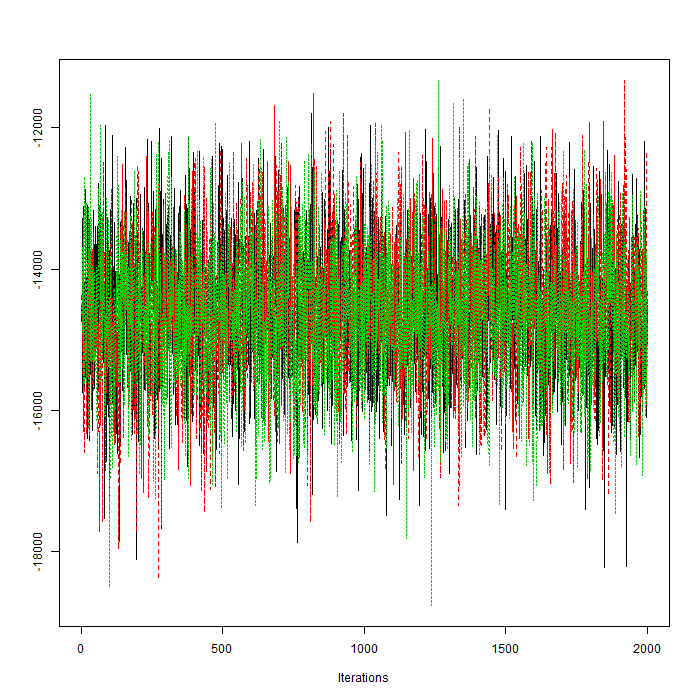} 
		\caption{} 
	\end{subfigure}
	\hfill
	\begin{subfigure}[t]{0.48\textwidth}
		\centering
		\includegraphics[width=\linewidth]{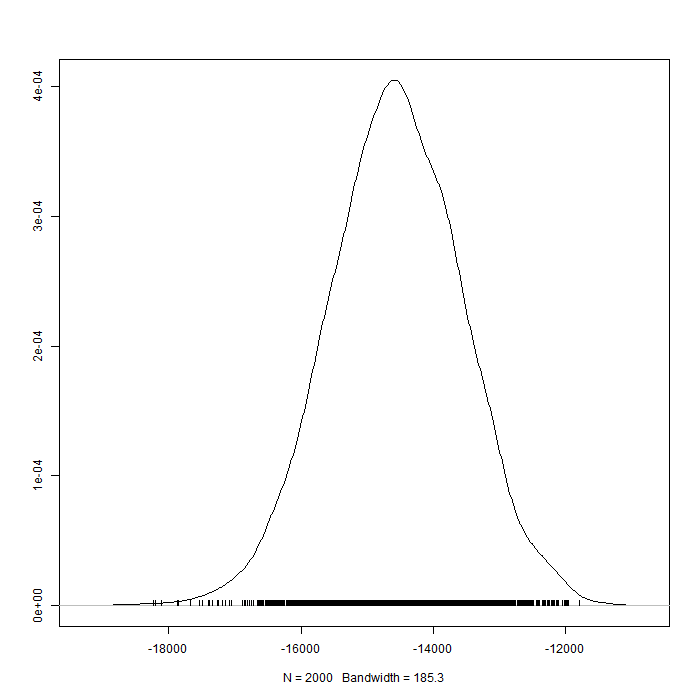} 
		\caption{} 
	\end{subfigure} \vfill
	\begin{subfigure}[t]{0.48\textwidth}
		\centering
		\includegraphics[width=\linewidth]{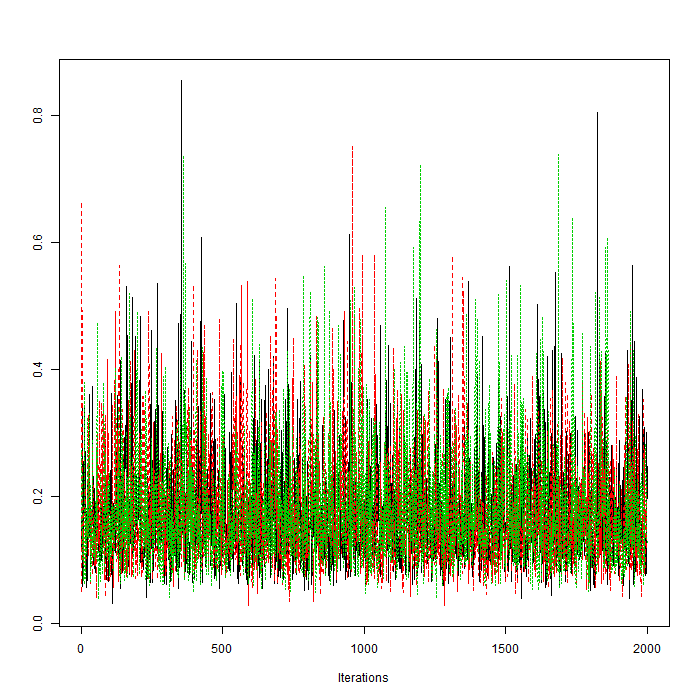} 
		\caption{} 
	\end{subfigure}
	\hfill
	\begin{subfigure}[t]{0.48\textwidth}
		\centering
		\includegraphics[width=\linewidth]{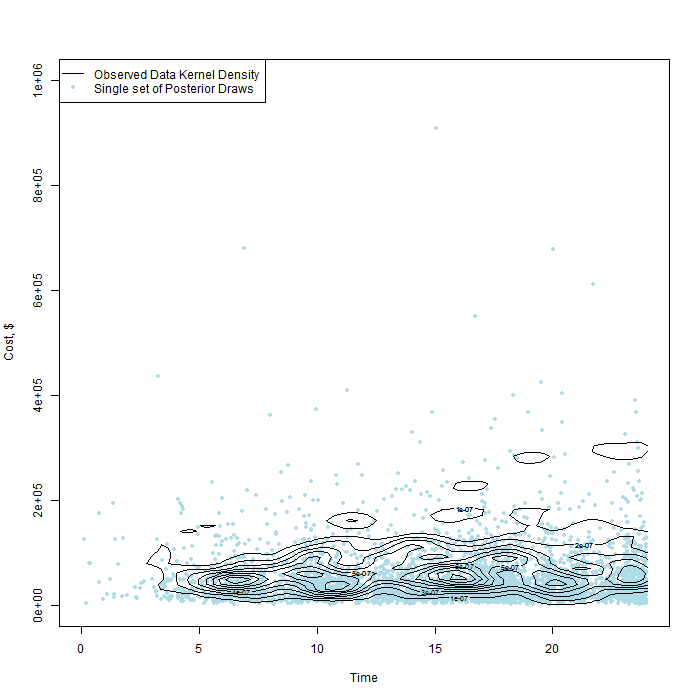} 
		\caption{} 
	\end{subfigure}
	\vspace{.1in}
	\caption{Diagnostic plots supporting data analysis results. Top row: traceplots of three MCMC chains of posterior NMB draws (left) and distribution of the combined posterior NMB draws of all chains (right). These NMB draws are based on $\kappa=\$50,000/12$. All three chains mix after starting with different initial clusters and seeds. Corresponding posterior is unimodal and peaked around $\$14,500$. Panel C shows the traceplots of three MCMC chains for DSI, which mix well. Finally, panel D shows a kernel density estimate of the joint observed time and cost distribution. In blue we show a single set of posterior predictive draws of joint cost and observed time. This shows adequate model fit: the posterior predictive is placing mass around the observed data. Moreover, the posterior predictive allows for occasional large cost draws. This indicates the local log-Normal cost distribution is able to capture skewness. If, for instance, the posterior predictive draws did not overlap with the observed data, we would be suspicious of the model fit.}
		\label{fig:diagnostic}

\end{figure}

We partition the interval from $[0, 24]$ into increments of $.5$. To sample from conditional posterior of $\{c_v \}_{1:V}$ (as mentioned Appendix \ref{ap:mcmc}) we use a Metropolis-Hastings update from jumping variance of $.5$. To sample from the posterior of $\theta$ (the covariate effects of the hazard model) we use a joint Metropolis-Hastings update with an initial identity covariance matrix multiplied by $.1$ along the diagonal. For both samplers, we adapt these jumping variances every 25 iterations starting from iteration 50 to iteration 200. Every $25^{th}$ iteration we use the previous 25 draws to target an acceptance rate of $23.4\%$, as per \cite{roberts2001}. Since we assume a log-normal cost distribution, posterior updates are conjugate using log-transformed cost. Figure \ref{fig:diagnostic} contains some diagnostic plots with a discussion in the caption. These plots show the MCMC chains to be well-mixed and model fit to be adequate. The total run-time was approximately 50 hours when parallelizing the three chains.

\begin{figure}
	\centering
	\includegraphics[scale=.5]{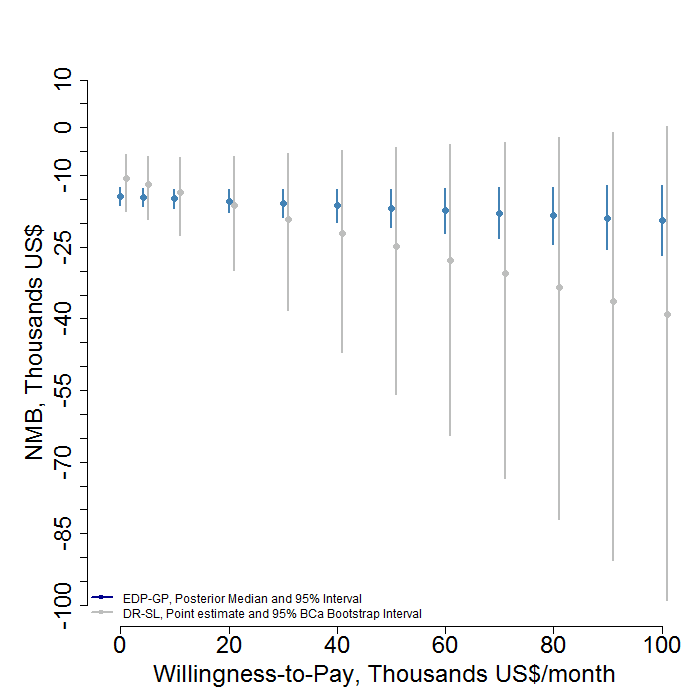} 
	\caption{NMB mean and 95\% bootstrap intervals for various willingness to pay from the DR-SL model in gray. The EDP-GP estimates from Figure \ref{fig:nmbest} are shown in blue for reference.}
	\label{fig:drslceac}
\end{figure}

For the doubly-robust (DR-SL) implementation of \cite{Li2018}, we estimate the propensity score model, cost model, and survival model using super learner with regression trees, GLM, and GLMnet as candidates. Inverse censoring probability weights were estimated using a discrete-time failure model described in Section 3.1.1 \cite{Li2018}. This is a logistic model that predicts the probability of censoring at each time point, conditional on not having been censored before that time point. The discretization is at the monthly level, thus there are 24 intervals in which one can be censored over $\tau=24$ months. The resulting model is used to predict the probability being censored at the observed time, for each subject. The inverse of this probability is the weight used in the DR approach. We include all Age, Household income, Charlson Index, and FIGO stage as covariates in each model. Due to small cell counts, we combined FIGO stage II and II-NOS into a single category. In the discrete-time failure model, we include a fixed effect for each month, 1-24. Due to sparsity, we included month as a continuous covariate rather than categorical in this model. In Figure \ref{fig:drslceac}, displays NMB estimates from this DR-SL model in gray, along with the EDP-GP estimates for reference. Note the larger uncertainty in the DR-SL model.

\end{appendix}
\end{document}

%% file: simulation_results.tex
\begin{table}
\centering
\caption{Simulation Results. Average bias of posterior mean NMB (as discussed in Section \ref{sc:causal} ) along with coverage and average width of 95\% credible/confidence interval (CI) is reported for EDP-GP model. Point estimate is reported for DR-SL along with coverage and width of 95\% bootstrap BCa interval. Bias is reported as a proportion of the truth. Censoring rate was 5\% in the low setting and 20\% in the high setting. Willingness-to-pay is set to $\kappa=1$. Results are across 200 simulated datasets with $N=1500$ subjects each.}
\label{tab:simres}
\begin{tabular}{cccccc|ccc}
\hline
\multicolumn{3}{c}{Simulation Setting}                                            & \multicolumn{3}{c}{EDP-GP}         & \multicolumn{3}{c}{DR-SL}       \\ \hline
True $Y$ Dist. & Joint Dist.                                     & \multicolumn{1}{l|}{Censoring} & Bias   & Coverage & Width & Bias  & Coverage & Width \\ \hline
\multicolumn{1}{l|}{\multirow{4}{*}{Gaussian}} & \multicolumn{1}{l|}{\multirow{2}{*}{Parametric}}& \multicolumn{1}{l|}{Low}      & -0.002 & 0.94    &  0.11  	    & -0.001 &   0.95  &  0.18          \\
\multicolumn{1}{l|}{}						  & \multicolumn{1}{l|}{}                           & \multicolumn{1}{l|}{High}      & -0.002 & 0.97    &  0.12         &  0.003 &   0.95  &  0.30          \\ 
\multicolumn{1}{l|}{}						  & \multicolumn{1}{l|}{\multirow{2}{*}{Bimodal}}   & \multicolumn{1}{l|}{Low}       & -0.01  & 0.94    &  0.13         &  0.11  &   0.60  &  0.64          \\
\multicolumn{1}{l|}{}						  & \multicolumn{1}{l|}{}                           & \multicolumn{1}{l|}{High}      & -0.01  & 0.94    &  0.14         &  0.16  &   0.40  &  0.77          \\ \hline
\multicolumn{1}{l|}{\multirow{4}{*}{Log-Normal}} & \multicolumn{1}{l|}{\multirow{2}{*}{Parametric}}& \multicolumn{1}{l|}{Low}    & -0.02   & 0.92    &  0.13  	    & -0.001 &   0.96  &  0.12          \\
\multicolumn{1}{l|}{}						  & \multicolumn{1}{l|}{}                           & \multicolumn{1}{l|}{High}      &  0.004   & 0.96    &  0.14         & -0.01  &   0.96  &  0.13          \\ 
\multicolumn{1}{l|}{}						  & \multicolumn{1}{l|}{\multirow{2}{*}{Bimodal}}   & \multicolumn{1}{l|}{Low}       & -0.004 & 0.98    &  0.11         &  0.02  &   0.94  &  0.18          \\
\multicolumn{1}{l|}{}						  & \multicolumn{1}{l|}{}                           & \multicolumn{1}{l|}{High}      &  0.03 	  & 0.92    &  0.12         &  0.06  &   0.90  &  0.20          \\ \hline

\end{tabular}
\end{table}

%% file: bsl_table.tex
\begin{table}[h!]
\centering
\caption{Sample Characteristics: Mean and sample standard deviations reported for continuous covariates. Counts and proportions reported for categorical covariates. Standardized mean differences (SMD) are provided. Typically $SMD>.1$ indicate large differences. Monetary amounts are in thousands of 2018 U.S. Dollars.}
\label{tbl:bsl}
\begin{tabular}{lccc} \hline
	\multicolumn{1}{l|}{}   &                          Radiation & 		Chemotherapy 	&    SMD       	\\
	\multicolumn{1}{l|}{}   &                             (N= 3,827 ) 	& 			(N= 245 ) 	&           	\\
	\hline
\multicolumn{1}{l|}{Total Accrued Costs (\$)}   	&  42.6 (36.8)  & 	51.3 (39.7)		&  .23        		\\
\multicolumn{1}{l|}{2-yr Survival Prob.}   		&   94.5       	& 	 93.0    		&          		\\
\multicolumn{1}{l|}{Age (years)}   					&   73.6 (6.2)  & 	73.2 (6.3)		&  .06	  		\\
\multicolumn{1}{l|}{Household Income (\$)}   		&   60.3 (28.8) & 	65.6 (34.0)		&  .17     		\\
\multicolumn{1}{l|}{Charlson Index }   				&          		&     				&  .12           \\
\multicolumn{1}{l|}{\ \ \ \ 0}   					&  	2176 (56.9) &     	131 (53.5)	&               \\
\multicolumn{1}{l|}{\ \ \ \ 1}   					&   1056 (27.6) &     	65 (26.5)	&               \\
\multicolumn{1}{l|}{\ \ \ \ 2}   					&   342 (8.9)   &     	30 (12.2)	&               \\
\multicolumn{1}{l|}{\ \ \ \ $\geq 3$}   			&   253 (6.6)   &     	19 (7.8)	&               \\
\multicolumn{1}{l|}{FIGO Stage}   					&    			&     				&   .5            \\
\multicolumn{1}{l|}{\ \ \ \ I-NOS}   				&   353 (9.2)   &     	23 (9.4)	&               \\
\multicolumn{1}{l|}{\ \ \ \ IA}   					&   1162 (30.4) &     	128 (52.2)	&               \\
\multicolumn{1}{l|}{\ \ \ \ IB}   					&   1780 (46.5)	&     	64 (26.1)	&               \\
\multicolumn{1}{l|}{\ \ \ \ II/II-NOS}  			&   532 (13.8)  &     	30 (12.2)	&               \\ \hline

\end{tabular}
\end{table}